\documentclass[sigconf]{acmart}
\usepackage{soul}
\usepackage{dsfont}
\usepackage{enumitem}

\renewcommand\footnotetextcopyrightpermission[1]{} % removes footnote with conference information in first column

\usepackage{booktabs} % For formal tables
\usepackage{mdframed}
\usepackage{framed}
\usepackage{array}
\usepackage{multirow}
\usepackage{pifont}
\newcommand{\tablemark}[1]{\makebox[1em][c]{\raisebox{0.4ex}[1.25ex][0.75ex]{\raisebox{-.5\height}{#1}}}}

\newcommand{\cmark}{\tablemark{\ding{51}}}
\newcommand{\xmark}{\tablemark{\ding{55}}}
\newcolumntype{C}[1]{>{\centering\arraybackslash}m{#1}}
\usepackage{hyphenat}
\usepackage{fnpct}
\usepackage{algpseudocode}
\usepackage{amsmath,epsfig}
\usepackage[boxed,ruled,vlined,linesnumbered]{algorithm2e}
\usepackage{amsthm}
\usepackage{setspace}
\usepackage{enumitem}
\usepackage{multirow}
\usepackage{hhline}
\usepackage{caption}
\usepackage{subcaption}
\usepackage{graphicx}
\usepackage{wrapfig}
\usepackage{amsmath,epsfig}
\usepackage{amsthm}

\newtheoremstyle{mystyle}%                % Name
  {}%                                     % Space above
  {}%                                     % Space below
  {\itshape}%                                     % Body font
  {}%                                     % Indent amount
  {\bfseries}%                            % Theorem head font
  {.}%                                    % Punctuation after theorem head
  { }%                                    % Space after theorem head, ' ', or \newline
  {}%                                     % Theorem head spec (can be left empty, meaning `normal')

\theoremstyle{mystyle}

\setcopyright{rightsretained}
\usepackage{tikz}
\usepackage{balance}

\usepackage{csquotes}
\usepackage{enumitem}
\usepackage{tablefootnote}
\usepackage{fnpct}
\usepackage{bbm}
\usepackage{lipsum}
\usepackage{setspace}

\makeatletter
\def\thm@space@setup{\thm@preskip=0pt
\thm@postskip=0pt}
\makeatother

\usepackage{array}
\usepackage{makecell}

\newcommand{\B}{\vspace*{-\smallskipamount}}

\newcommand{\BBB}{\vspace*{-\bigskipamount}}

\usepackage{longtable}
\usepackage{mdframed}
\usepackage{colortbl}
\usepackage[hang,flushmargin]{footmisc}
\usepackage{natbib}
\newcommand*\wrapletters[1]{\wr@pletters#1\@nil}
\def\wr@pletters#1#2\@nil{#1\allowbreak\if&#2&\else\wr@pletters#2\@nil\fi}
\usepackage{tikz}
\usetikzlibrary{automata,matrix,shapes,arrows,positioning,chains,calc}
\usetikzlibrary{snakes}
\usetikzlibrary{arrows,scopes}
\usetikzlibrary{positioning,chains,fit,shapes,calc}
\usepackage{caption}
\usepackage{subcaption}
\usepackage{amsmath}
\usepackage{lineno}

\definecolor{lightyellow}{HTML}{FFFFFF}
\sethlcolor{lightyellow}
\definecolor{myblue}{HTML}%{6F00FF}
{BF40BF}
\mdfdefinestyle{MyFrame}{%
    linecolor=white,
    outerlinewidth=2pt,
    innertopmargin=4pt,
    innerbottommargin=4pt,
    innerrightmargin=4pt,
    innerleftmargin=4pt,
        leftmargin = 4pt,
        rightmargin = 4pt,
        backgroundcolor=yellow!07!white}

        \usepackage{array}
\usepackage{multirow}
\usepackage{pifont}
\newcolumntype{C}[1]{>{\centering\arraybackslash}p{#1}}
\usepackage{array}
\usepackage{makecell}
\usepackage{array}
\usepackage{graphicx}
\usepackage{bbold}
\newcolumntype{C}[1]{>{\centering\arraybackslash}m{#1}}

\usetikzlibrary{arrows.meta,positioning,calc}

\definecolor{lightgreenbox}{RGB}{226,245,226}

\DeclareRobustCommand{\algname}{\textsc{DexterSQL}}

\begin{document}

% %==============================================================
% %reduce space before and after equations
% \setlength{\belowdisplayskip}{0pt}
% \setlength{\belowdisplayshortskip}{0pt}
% \setlength{\abovedisplayskip}{0pt}
% \setlength{\abovedisplayshortskip}{0pt}
% %==============================================================

\onecolumn
\fancyhead{}
 \title{\textsc{DexterSQL}: \underline{D}eep Schema \underline{Ex}ploration and Rule-based Correction for \underline{T}ext-to-SQL Gen\underline{er}ation}

\author{
Anik Pramanik$^{1}$, Murat Kantarcioglu$^{2}$, Vincent Oria$^{1}$, Shantanu Sharma$^{1}$
}
\affiliation{%
\institution{$^{1}$New Jersey Institute of Technology, USA. \quad $^{2}$Virginia Tech, USA.}
\country{}}

%==============================================================
%reduce space before and after equations
\setlength{\belowdisplayskip}{0pt}
\setlength{\belowdisplayshortskip}{0pt}
\setlength{\abovedisplayskip}{0pt}
\setlength{\abovedisplayshortskip}{0pt}
%==============================================================

\begin{abstract}
Prompting-based (\textit{i}.\textit{e}., non-fine-tuning) Text-to-SQL methods, where underlying large language model parameters are not changed for the task, face three problems: 
(\textit{i})~relying on coarse-grained schema information that may not reveal the fine-grained relationships needed to distinguish ambiguous columns, 
(\textit{ii})~failing to capture recurring SQL-generation failures, and 
(\textit{iii})~suffering from omission or hallucination of components in complex questions.

This paper develops \textsc{DexterSQL}, a prompting/non-fine-tuning-based Text-to-SQL system that improves SQL generation with three novel components: 
(\textit{i})~\emph{deep schema explorator} that identifies ambiguous columns, analyzes their individual and joint data distributions to uncover their relationships and the distinct role of each,  
(\textit{ii})~\emph{database-agnostic rule creator} that mines mismatches between generated and gold SQL only on the training database and converts them into database-agnostic corrective rules that capture recurring LLM failure patterns; and 
(\textit{iii})~\emph{multi-path SQL generation} that introduces a dependency-tree-based intermediate representation that uses the question's sentence structure to guide its decomposition into an SQL skeleton for final SQL generation.

\textsc{DexterSQL} achieves a higher accuracy compared to the state-of-the-art using both open-source/weight and closed-source/weight models. 
Particularly, \textsc{DexterSQL} shows a high improvement of at least 5.5\% using an open-weight model (GPT-OSS-120B) on BIRD-Dev, with total accuracy 70.4\%.
\textsc{DexterSQL} also shows better improvement of at least 1.4\% using closed-weight models, with total accuracy 72.1\% and 72.9\% on BIRD-Dev with GPT-4o and GPT-5.2.

\end{abstract}

 \maketitle

\section{Introduction}
\label{sec:intro}

Text-to-SQL approaches translate a natural-language (NL) question into an executable SQL query for execution on a relational database. 
% allowing non-expert users to access data without knowing SQL or the underlying schema~\cite{li2023can}. 
%The problem has a long line of work. 
Early Text-to-SQL systems were based on task-specific neural architectures, \textit{e}.\textit{g}., encoder-decoder models~\cite{zhong2017seq2sql}, grammar- and sketch-based decoders~\cite{xu2017sqlnet}, and schema-aware encoders~\cite{wang2020acl}. 
These systems were trained separately for each dataset~\cite{zhong2017seq2sql,guo2019acl,wang2020acl,gan2021natural}, and thus are less robust to unseen schemas, lack a general understanding of NL, and do not generalize on different datasets.  
Recently, large language models (LLMs) have substantially advanced Text-to-SQL process and surpassed these specialized models on standard benchmarks such as Spider~\cite{yu2018spider} and BIRD~\cite{li2023can}.  

\smallskip
\noindent
\textbf{LLM-based Text-to-SQL pipeline.}
%Rather than emitting SQL in a single step, 
LLM-based Text-to-SQL systems/approaches~\cite{cao2024rsl,nahid2026rethinking,pourreza2023din,wang2025mac,li2026deepeye,xie2025opensearch,pourreza2024chase,gao2023text} decompose the task into a sequence of steps, such as \emph{schema linking} that narrows the full database schema to the tables and columns relevant to a NL question; %~\cite{cao2024rsl,nahid2026rethinking}; 
\emph{SQL generation} that produces one or more candidate SQL queries;  %, often by decomposing the question or reasoning over its structure~
%\cite{pourreza2023din,wang2025mac}; 
 \emph{correction} that repairs invalid or inconsistent candidate SQL queries;  % that are , sometimes using execution feedback~ 
 %\cite{li2026deepeye,xie2025opensearch}; 
 and \emph{selection} that selects a final SQL query from the candidate pool. % \cite{pourreza2024chase,gao2023text}.

\smallskip\noindent\textbf{Fine-tuning vs.\ non-fine-tuning.}
LLM-based Text-to-SQL approach can be broadly categorized into two classes. \emph{Fine-tuning (FT)} methods modifies an LLM's weights on Text-to-SQL data by either training the generator directly or through auxiliary supervision~\cite{pourreza2024dts,li2024codes,li2025omnisql,yang2024acl,liu2026xiyan}. \emph{Non-fine-tuning (non-FT)} methods keep the LLM frozen and steer it purely through prompting, schema linking, NL decomposition, and  correction~\cite{pourreza2023din,gao2023text,xie2025opensearch,li2026deepeye,shkapenyuk2025automatic,wang2025mac, cao2026apex, somayajula2026soma,cao2024rsl,dong2023c3,lee-etal-2025-mcs, chung2025long}.

The two classes tradeoff differently. Fine-tuning can inject dataset- and schema-specific knowledge directly into the model to produce more accurate, self-contained models. However, this requires large amounts of labeled training data and must be repeated for every model or target domain. 
%and the open models that can be fine-tuned still lag behind the strongest proprietary LLMs. 
In contrast, non-fine-tuning methods require no training: they only manage the context and prompt provided to an off-the-shelf LLM. 
This makes them %inexpensive to use and 
to improve automatically as base models advance, though their accuracy is bounded by the underlying LLM and is sensitive to the context they are given. 
These benefits, especially no training requirements, make non-fine-tuning methods easy for anyone to deploy. 

%\smallskip\noindent\textbf{Why we focus on non-fine-tuning.}
 \smallskip%\noindent\textbf{Our approach.}
\noindent\textbf{\algname{}.} This paper 
\emph{\textbf{focuses on the non-fine-tuning setting}}, due to the benefits offered by them and ease of deployment.  
%because it is the more broadly usable and durable one. 
%Fine-tuning is out of reach for many practitioners: it demands labeled corpora and expensive compute that most do not have, and its gains must be re-earned each time a stronger base model appears. In contrast, 
%
%and inherit improvements in the underlying models for free. Making these training-free pipelines more accurate therefore has broad and lasting impact.
%
%\smallskip
%\noindent\textbf{\algname{}.} %Challenges.}
However, despite rapid progress, % ~\cite{pourreza2023din,gao2023text,xie2025opensearch,li2026deepeye,shkapenyuk2025automatic,wang2025mac, cao2026apex, somayajula2026soma,cao2024rsl,dong2023c3,lee-etal-2025-mcs, chung2025long}, 
such Text-to-SQL systems still face three major  challenges.

\emph{(C1)~Detecting underlying database insights.} To produce a SQL query, LLM needs to understand the context from either the schema and/or sampled values (as {\color{black}in~\cite{shkapenyuk2025automatic,li2026deepeye,xie2025opensearch}}). 
However, such information may not be sufficient to distinguish ambiguous/confusable columns, whose roles become clear only by analyzing their data distributions and relationships across  tables.
%Existing systems explore the database only at coarse-grained level, leaving ambiguous columns unresolved~\cite{shkapenyuk2025automatic,xie2025opensearch,li2026deepeye}.

\emph{(C2)~Capturing recurring model errors.} A Text-to-SQL system can have recurring %, database-agnostic 
SQL-generation failures due to the underlying LLM performance, so these reusable failure patterns must be separated from database-specific mistakes and converted into guidance for correcting future queries.

\emph{(C3)~effective SQL generation.} LLM-based SQL generation systems may omit, hallucinate, or misplace conditions and relationships in a complex NL question, producing executable SQL that answers an incomplete or different question.

\definecolor{DarkGreen}{RGB}{0,100,0}

\begin{table*}[t]
\BBB
\centering
\small
\renewcommand{\arraystretch}{1}
\resizebox{\textwidth}{!}{%
\begin{tabular}{|p{2.2cm}|C{1.3cm}|C{1.2cm}|C{1.7cm}|C{2.1cm}|C{1.1cm}|C{1.5cm}|C{1.6cm}|C{1.6cm}|C{1.5cm}|C{1.5cm}|C{1.6cm}|C{1.0cm}|C{1.0cm}|}
\hline
\multirow{2}{*}{\textbf{Method}} &
\multicolumn{4}{c|}{\textbf{Pre-processing} ({\color{blue}\S\ref{sec:preprocessing}})} &
\multirow{2}{1.4cm}{\centering\textbf{Schema Linking} ({\color{blue}\S\ref{sec:schema-linking}})} &
\multicolumn{4}{c|}{\textbf{Text-to-SQL Generation} ({\color{blue}\S\ref{sec:sql-generation}})} &
\multicolumn{2}{c|}{\textbf{Post-processing} ({\color{blue}\S\ref{sec:sql-correction-selection}})} &
\multicolumn{2}{c|}{\textbf{Accuracy}} \\
\cline{2-5}\cline{7-14}
&
\textbf{DB Value Matching ({\color{blue}\S\ref{sec:indexing}})} &
\textbf{Additional Context ({\color{blue}\S\ref{sec:profiling}})} &
\textbf{Deep Schema Exploration ({\color{blue}\S\ref{sec:deep-exploration}})} &
\textbf{Database-Agnostic Rule Creation ({\color{blue}\S\ref{sec:rule-creation}})} &
&
\textbf{Ambiguity Resolver ({\color{blue}\S\ref{sec:note-injection}})} &
\textbf{Dependency Tree ({\color{blue}\S\ref{sec:sql-generator}})} &
\textbf{Divide-and-Conquer ({\color{blue}\S\ref{sec:sql-generator}})} &
\textbf{Few-shot Examples ({\color{blue}\S\ref{sec:sql-generator}})} &
\textbf{Consistency ({\color{blue}\S\ref{sec:selection}})} &
\textbf{Rule-Based Correction ({\color{blue}\S\ref{sec:correction}})} &
\textbf{Spider-Test} &
\textbf{BIRD-Dev} \\
\hline
DAIL-SQL~\cite{gao2023text}
& {\color{red}\xmark} & {\color{red}\xmark} & {\color{red}\xmark} & {\color{red}\xmark}
& {\color{red}\xmark} & {\color{red}\xmark}
& {\color{red}\xmark} & {\color{red}\xmark} & {\textcolor{DarkGreen}\cmark}
& {\textcolor{DarkGreen}\cmark} & {\color{red}\xmark}
& 74.0 & 51.8 \\
\hline
C3~\cite{dong2023c3}
& {\color{red}\xmark} & {\color{red}\xmark} & {\color{red}\xmark} & {\color{red}\xmark}
& {\textcolor{DarkGreen}\cmark} & {\color{red}\xmark}
& {\color{red}\xmark} & {\color{red}\xmark} & {\textcolor{DarkGreen}\cmark}
& {\textcolor{DarkGreen}\cmark} & {\color{red}\xmark}
& 69.3 & 53.8 \\
\hline
Rethinking~\cite{nahid2026rethinking}
& {\textcolor{DarkGreen}\cmark} & {\textcolor{DarkGreen}\cmark} & {\color{red}\xmark} & {\color{red}\xmark}
& {\textcolor{DarkGreen}\cmark} & {\color{red}\xmark}
& {\color{red}\xmark} & {\color{red}\xmark} & {\color{red}\xmark}
& {\color{red}\xmark} & {\color{red}\xmark}
& 72.2 & 54.6 \\
\hline
DIN-SQL~\cite{pourreza2023din}
& {\color{red}\xmark} & {\color{red}\xmark} & {\color{red}\xmark} & {\color{red}\xmark}
& {\textcolor{DarkGreen}\cmark} & {\color{red}\xmark}
& {\color{red}\xmark} & {\textcolor{DarkGreen}\cmark} & {\textcolor{DarkGreen}\cmark}
& {\color{red}\xmark} & {\color{red}\xmark}
& 70.0 & 56.2 \\
\hline
AutoLink~\cite{wang2026autolink}
& {\textcolor{DarkGreen}\cmark} & {\textcolor{DarkGreen}\cmark} & {\color{red}\xmark} & {\color{red}\xmark}
& {\textcolor{DarkGreen}\cmark} & {\color{red}\xmark}
& {\color{red}\xmark} & {\color{red}\xmark} & {\color{red}\xmark}
& {\color{red}\xmark} & {\color{red}\xmark}
& 77.5 & 57.4 \\
\hline
OpenSearchSQL~\cite{xie2025opensearch}
& {\textcolor{DarkGreen}\cmark} & {\color{red}\xmark} & {\color{red}\xmark} & {\color{red}\xmark}
& {\textcolor{DarkGreen}\cmark} & {\color{red}\xmark}
& {\color{red}\xmark} & {\color{red}\xmark} & {\textcolor{DarkGreen}\cmark}
& {\textcolor{DarkGreen}\cmark} & {\color{red}\xmark}
& 67.8 & 58.2 \\
\hline
RSL-SQL~\cite{cao2024rsl}
& {\textcolor{DarkGreen}\cmark} & {\color{red}\xmark} & {\color{red}\xmark} & {\color{red}\xmark}
& {\textcolor{DarkGreen}\cmark} & {\color{red}\xmark}
& {\color{red}\xmark} & {\color{red}\xmark} & {\color{red}\xmark}
& {\textcolor{DarkGreen}\cmark} & {\color{red}\xmark}
& 74.6 & 59.3 \\
\hline
Alpha-SQL~\cite{li2025alpha}
& {\textcolor{DarkGreen}\cmark} & {\color{red}\xmark} & {\color{red}\xmark} & {\color{red}\xmark}
& {\textcolor{DarkGreen}\cmark} & {\color{red}\xmark}
& {\color{red}\xmark} & {\textcolor{DarkGreen}\cmark} & {\color{red}\xmark}
& {\textcolor{DarkGreen}\cmark} & {\color{red}\xmark}
& 80.3 & 62.8 \\
\hline
ApexSQL~\cite{cao2026apex}
& {\textcolor{DarkGreen}\cmark} & {\textcolor{DarkGreen}\cmark} & {\color{red}\xmark} & {\color{red}\xmark}
& {\textcolor{DarkGreen}\cmark} & {\color{red}\xmark}
& {\color{red}\xmark} & {\textcolor{DarkGreen}\cmark} & {\color{red}\xmark}
& {\textcolor{DarkGreen}\cmark} & {\color{red}\xmark}
& 79.1 & 64.2 \\
\hline
DeepEye-SQL~\cite{li2026deepeye}
& {\textcolor{DarkGreen}\cmark} & {\color{red}\xmark} & {\color{red}\xmark} & {\color{red}\xmark}
& {\textcolor{DarkGreen}\cmark} & {\color{red}\xmark}
& {\color{red}\xmark} & {\textcolor{DarkGreen}\cmark} & {\textcolor{DarkGreen}\cmark}
& {\textcolor{DarkGreen}\cmark} & {\color{red}\xmark}
& 81.9 & 64.9 \\
\hline

{\cellcolor[HTML]{C1FD8E}{\textbf{\textsc{DexterSQL}}}}
& 

 {\cellcolor[HTML]{C1FD8E}{\textbf{{\textcolor{DarkGreen}\cmark}}}} 
 &  {\cellcolor[HTML]{C1FD8E}{\textbf{{\textcolor{DarkGreen}\cmark}}}} 

 & {\cellcolor[HTML]{C1FD8E}{\textbf{{\textcolor{DarkGreen}\cmark}}}} 
 & {\cellcolor[HTML]{C1FD8E}{\textbf{{\textcolor{DarkGreen}\cmark}}}}
 & {\cellcolor[HTML]{C1FD8E}{\textbf{{\textcolor{DarkGreen}\cmark}}}} &  
 {\cellcolor[HTML]{C1FD8E}{\textbf{{\textcolor{DarkGreen}\cmark}}}}
  & {\cellcolor[HTML]{C1FD8E}{\textbf{{\textcolor{DarkGreen}\cmark}}}} & 
 
  {\cellcolor[HTML]{C1FD8E}{\textbf{{\textcolor{DarkGreen}\cmark}}}} 
  & 
  {\cellcolor[HTML]{C1FD8E}{\textbf{{\textcolor{DarkGreen}{\textcolor{DarkGreen}\cmark}}}}}

  & 
  {\cellcolor[HTML]{C1FD8E}{\textbf{{\textcolor{DarkGreen}\cmark}}}} 

&  {\cellcolor[HTML]{C1FD8E}{\textbf{{\textcolor{DarkGreen}\cmark}}}}

& {\cellcolor[HTML]{C1FD8E}{\textbf{85.6}}} & {\cellcolor[HTML]{C1FD8E}{\textbf{70.4}}} \\

\hline
\end{tabular}%
}

\caption{
Comparison of non-fine-tuning Text-to-SQL methods.
{\scriptsize \textnormal{Notations: \textit{DB Value Matching}: looking up or embedding database cell values for grounding; \textit{Additional Context}: extracting column statistics or metadata; \textit{Deep Schema Exploration}: investigating data distributions and relationships between ambiguous columns; \textit{Database-Agnostic Rule Creation}: mining recurring database-agnostic SQL-generation failures into reusable rules; \textit{Schema Linking}: selecting the tables and columns relevant to a question; \textit{Ambiguity Resolver}: selecting and incorporating guidance that distinguishes confusable columns; \textit{Dependency Tree}: using dependency structure to guide decomposition into an intermediate SQL skeleton; \textit{Divide-and-Conquer}: breaking a question into simpler sub-questions; \textit{Few-shot Examples}: using retrieved examples as demonstrations; \textit{Consistency}: selecting among multiple candidates through voting or agreement; and \textit{Rule-Based Correction}: applying synthesized rules to detect and correct recurring SQL-generation failures.}}}
\label{tab:non_ft_comparison1}
\BBB\BBB\BBB\B
\end{table*}

% % \smallskip\noindent\textbf{Our approach.}
% \noindent\textbf{\algname{}.} 
We develop \textbf{\algname{}}, a non-fine-tuning Text-to-SQL system that addresses the above-mentioned challenges. 
\algname{} introduces three novel components, each addressing the above challenges:
(\textit{i})~\emph{deep schema explorator}, an offline (\textit{i}.\textit{e}., before arrival of any NL query) analysis to mine data-level relationships among ambiguous  columns into compact disambiguation notes; 
(\textit{ii})~\emph{database-agnostic rule creator}, an offline process to learn recurring model failures using \emph{\textbf{only}} the training dataset\footnote{\scriptsize NL queries are asked on a database, entitled \emph{\textbf{target database}} that   \textbf{is completely disjoint from the training database}.} and build database-agnostic correction rules that help for future NL questions; 
and 
(\textit{iii})~\emph{multi-path SQL generation}, which introduces a dependency-tree-based intermediate representation to preserve question elements and combines it with few-shot in-context-learning and divide-and-conquer generation to produce multiple SQL candidates.

%With frontier models, 

%As shown in  

%Unlike prior non-fine-tuning systems, \algname{} derives its guidance from the database contents and from observed model behavior rather than from surface-level schema and example similarity, and it applies this guidance without any model fine-tuning

% \algname{} introduces three new components, each designed to address a key challenge faced by non-fine-tuning Text-to-SQL systems.
% The three new components increase the accuracy of Text-to-SQL. 
% Particularly, 
% \algname{} achieves 67.6\% on BIRD-Dev with GPT-OSS-120B (a public LLM), improving over current non-fine-tuning methods (by at least {\color{red}??\%.}); see the last two columns of {\color{blue}Table~\ref{tab:bird-sql-generation}}.
% Also, \algname{} achieves 
% 71.6\% on BIRD-Dev with GPT-4o and 72.2\% with GPT-5.2, and this is at least {\color{red}??\%} more compared to the current state-of-the-art~\cite{li2026deepeye,apexsql paper}.

\emph{\textbf{\algname{} Performance.}} 
Text-to-SQL systems can be deployed using either open-weight, free LLMs (\textit{e}.\textit{g}., GPT-OSS-120B) or proprietary (closed-weight) LLMs (\textit{e}.\textit{g}., GPT-4o and GPT-5.2). These two classes of models show different tradeoffs. %Open-weight models can be deployed locally, eliminating the need to transmit database schemas or other potentially sensitive information to third-party services. Thus, they are the best for privacy-sensitive applications.
Open-weight models can be deployed within an organization’s local or private infrastructure, eliminating the need to transmit database schemas or other potentially sensitive information to third-party services. At sufficient scale, local deployment of smaller LLM models can also reduce marginal inference costs, and operational security and privacy.
~ 
However, their SQL generation accuracy generally lags behind that of state-of-the-art proprietary LLMs. %In contrast, proprietary models typically achieve higher accuracy but require access to the database schema, metadata, and/or content and deployment cost. 
In contrast, proprietary models often provide stronger out-of-the-box performance, but they may require transmitting database schemas, metadata, or content to third-party services and can incur higher deployment and usage costs.

%\algname{} not only outperforms existing work on proprietary models, but also achieves much higher accuracy on open-weight models, making it well-suited for sensitive data.
\algname{} not only outperforms prior methods when using proprietary models, but also achieves substantially higher accuracy with open-weight models, making it particularly well suited for sensitive-data applications and local or cost-efficient deployment.
Particularly, 
on BIRD-Dev benchmark~\cite{li2023can}, 
\algname{} achieves an execution accuracy, EX (\textit{i}.\textit{e}., percentage of queries whose execution results match the gold-SQL results), of 70.4\% using GPT-OSS-120B, outperforming existing non-fine-tuned Text-to-SQL systems by at least 5.5\%; see the last column of~{\color{blue}Table~\ref{tab:non_ft_comparison1}}. 
Also, \algname{} achieves EX of 72.1\% with GPT-4o and 72.9\% with GPT-5.2, exceeding the state-of-the-art non-fine-tuning methods by at least 1.4\%~\cite{li2026deepeye,cao2026apex}.

\begin{mdframed}[
  linewidth=0.6pt,
  backgroundcolor=yellow!8,
  innerleftmargin=2pt,
  innerrightmargin=2pt,
  innertopmargin=2pt,
  innerbottommargin=2pt
]
\noindent\textbf{\algname{} code are given at: \\
https://github.com/SecretDeB/DexterSQL}
\end{mdframed}

\section{\algname{}}

\algname{} develops three novel components, each designed to tackle one of the key limitations of non-fine-tuned Text-to-SQL systems. Together, these components substantially improve SQL generation accuracy.
~
This section provides contributions and high-level overview of \algname{}.

\subsection{Challenges and \algname{}}
\label{subsection:challenges_and_our_solutions}

To better understand what \algname{} brings in, we  organize our contributions around the above-mentioned three challenges and illustrate them using two example tables; see 
{\color{blue}Tables~\ref{tab:running-patient}} and {\color{blue}\ref{tab:running-examination}}, which are taken from the \textsf{thrombosis\_prediction} database in BIRD~\cite{li2023can}. 
%: \textsf{Patient}(\underline{\textsf{ID}}, \textsf{Diagnosis}, $\ldots$) and \textsf{Examination}(\textsf{ID}, \textsf{Diagnosis}, $\ldots$), shown in . The displayed rows are illustrative, and the omitted columns are not needed for the running example. Both tables contain a \textsf{Diagnosis} column with disease names such as \textsf{SLE}, and \textsf{PSS}. However, 
\textsf{Patient.Diagnosis} stores the patient's final diagnosis, whereas \textsf{Examination.Diagnosis} stores a preliminary diagnosis recorded for each examination. We use these tables to explain the three challenges and the corresponding \algname{} components.

\begin{table}[!h]
%\BBB
%\scriptsize
\centering
%\captionsetup{font=footnotesize}
\begin{minipage}[t]{0.4\columnwidth}
\centering
\begin{tabular}{|c|c|} \hline
\underline{\textsf{ID}} & \textsf{Diagnosis} \\\hline
1 & SLE \\
2 & PSS \\
\hline
\end{tabular}
\captionof{table}{\textsf{Patient} table.}
\label{tab:running-patient}
\end{minipage}
\hfill
\begin{minipage}[t]{0.45\columnwidth}
\centering
\begin{tabular}
%{\linewidth}{@{\extracolsep{\fill}}
{|c|c|}\hline
\textsf{ID} & \textsf{Diagnosis} \\\hline
1 & SLE \\
1 & PSS \\
2 & SLE \\\hline
\end{tabular}
\captionof{table}{\textsf{Examination} table.}
\label{tab:running-examination}
\end{minipage}
%\BBB%\BBB%\BBB

\end{table}

\medskip
\noindent\textbf{Challenge 1: Detecting underlying database insights.} 
The first challenge is for an LLM to correctly select the desired tables and columns of a given database when an NL query is posted. 

% to the system.

% understand fine-grained details of the database (\textit{e}.\textit{g}.,  relationships between columns --- analyzing how a column relates to its table or other tables through the values it contains) to overcome ambiguity in understanding the schema.
%Many Text-to-SQL errors require this deeper database exploration to resolve ambiguity and generate the correct SQL query.

%To understand the challenge, 
For intance, consider a question: ``\emph{what is the final diagnosis of the patient with \textsf{ID}~1?}'' Answering it requires reading a \textsf{Diagnosis} column, but a surface-level (\textit{i}.\textit{e}., coarse-grain--- 
examining only individual column values or relying only on schema) understanding of the schema cannot determine which \textsf{Diagnosis} column (\textsf{Patient.Diagnosis} or \textsf{Examination.Diagnosis}) should be used.

This example shows LLMs need to understand fine-grained details of the database (\textit{e}.\textit{g}.,  relationships between columns --- analyzing how a column relates to its table or other tables through the values it contains) to overcome ambiguity in understanding the schema. 
~
This distinction is not obvious from the column names or from the overlapping values and can only be clear from the underlying data distribution: \textsf{Patient.Diagnosis} has one row per patient, whereas \textsf{Examination.\allowbreak Diagnosis} can have multiple rows for the same \textsf{Patient.ID}. 
Since \textsf{Patient.Diagnosis} records a single, final diagnosis per patient, it is the correct column for this patient-level question---a choice that requires inspecting the data distribution, not just the schema.

Existing prompt/non-fine-tuning-based approaches~\cite{li2026deepeye, shkapenyuk2025automatic,pourreza2023din,gao2023text,cao2026apex,cao2024rsl,xie2025opensearch, dong2023c3,wang2026autolink, nahid2026rethinking,lee-etal-2025-mcs}, however, explore databases at a surface level.

\medskip
\noindent
\textbf{Our solution: Deep Schema Explorator ({\color{blue}\S\ref{sec:deep-exploration}}).}
\algname{} %uses LLM judges to 
identifies column pairs that are genuinely ambiguous for SQL generation. For each ambiguous pair, it examines the distribution within each column and  joint distribution 
%, including its row count, missing values, distinct values, and frequent values, and 
to 
measures how much the two columns' value sets overlap. 
~
When their tables can be joined, \algname{} additionally determines the fraction of records participating in the relationship
%, whether a record on one side corresponds to multiple records on the other, 
and how often the ambiguous values agree across joined records. An LLM summarizes this information into a concise disambiguation note explaining how the columns differ and when each should be used.

For example, {\color{blue}Table~\ref{tab:running-patient}} contains one diagnosis for patient~1, whereas {\color{blue}Table~\ref{tab:running-examination}} contains two examination diagnoses for the same patient. After joining on \textsf{ID}, one patient record therefore corresponds to multiple examination records, and the diagnosis values do not always agree: the patient diagnosis is \textsf{SLE}, while the examination diagnoses are \textsf{SLE} and \textsf{PSS}. Given this information, the LLM creates a note stating that \textsf{Patient.Diagnosis} represents the patient-level final diagnosis and should be used for final-diagnosis questions, whereas \textsf{Examination.Diagnosis} represents diagnoses associated with individual examination events.

\medskip
\noindent
\textbf{Challenge 2: Capturing recurring model errors. }% and dataset preferences.} % and recurring LLM failures.}
The second challenge is to capture recurring SQL-generation failures of the underlying LLM. %, as well as dataset-specific SQL preferences when they affect the expected answer. 

For instance, consider 
%Using the schema of running example, %suppose \textsf{Examination.Diagnosis}  
a question: ``\emph{What is the ratio of examinations diagnosed as \textsf{PSS} to examinations diagnosed as \textsf{SLE}?}'' The intended ratio is $1/2=0.5$. However, an LLM may generate \textsf{SUM(Diagnosis = `PSS') / SUM(Diagnosis = `SLE')}, which may return 0 due to integer division. The correct formulation needs to produces a floating output via \textsf{CAST(SUM(Diagnosis = 'PSS') AS REAL) / SUM(Diagnosis = 'SLE')}, which returns $0.5$. 

This example shows that while an LLM generates an executable SQL that uses the correct tables and columns, but still returns the wrong result due to missing a SQL-formulation detail, \textit{e}.\textit{g}.,  integer versus floating-point division. Avoiding this error requires the LLM to recognize that a ratio needs floating-point division and insert a type conversion. %, such as \textsf{CAST(SUM(Diagnosis = 'PSS') AS REAL) / SUM(Diagnosis = 'SLE')}.

Existing prompt-based approaches use few-shot example selection~\cite{shkapenyuk2025automatic,li2026deepeye,gao2023text, dong2023c3, xie2025opensearch} that select examples based on overall NL-question similarity, without isolating the phrases, SQL operators, or query constructs responsible for a recurring error. Hence, a retrieved example may appear relevant at a high level while lacking the specific SQL formulation needed to avoid the failure. 

% SQL-correction approaches, on the other hand, often rely on a small set of generic, manually defined rules. Such fixed rules may not capture the distinct recurring failure patterns exhibited by different LLMs.

%Prompt-based approaches that perform dedicated  often rely on a small set of generic, manually defined rules. However, different LLMs may exhibit different generation biases that produce different types of LLM failures, and different datasets may have different preferences in SQL.

%and select examples based on overall NL question similarity or preliminary SQL similarity, without isolating the NL phrases, SQL operators, or query constructs that cause the recurring error. {\color{black} 
% As a result, they may retrieve examples that appear similar at a high level but do not contain the specific SQL structure needed to avoid the failure. 
% Prompt-based approaches that perform dedicated post-generation SQL correction, such as~\cite{gong2026sqlens}, often rely on a small set of generic, manually defined rules. However, different LLMs may exhibit different generation biases that produce different types of LLM failures, and different datasets may have different preferences in SQL.} 

\medskip
\noindent
\textbf{Our solution: Rule Creation ({\color{blue}\S\ref{sec:rule-creation}}).}
\algname{} uses \emph{\textbf{training data}}\footnote{\scriptsize A training data contains NLQ and gold-SQL pair and is \textbf{disjoint from testing data on which NL queries will be asked.}} to mine recurring SQL-generation failures and convert them into database-agnostic rules for correcting generated SQL. 
First, it generates candidate SQL for sampled training questions and retains candidates whose results do not match the corresponding gold SQL. 
It then uses an LLM to explain the cause of each failure and removes explanations tied to a particular database, such as incorrect schema linking or misunderstanding a specific column. Next, it clusters the remaining database-agnostic explanations within and across training databases so that failures with the same underlying cause are grouped together. Finally, it uses each recurring failure group to create a \emph{\textbf{database-agnostic correction rule}} that can be applied after SQL generation.

For the ratio example, the generated SQL returns $0$ instead of the intended $0.5$ because it divides two integer-valued aggregates without a floating-point conversion. From similar failures in the training data, \algname{} creates a rule instructing the correction stage to cast one operand to a floating-point type before division. When this failure pattern appears in a generated query, \algname{} selects the rule and applies its correction.

\begin{table}[!t]
\scriptsize
    \centering
    \begin{tabular}{|p{1.9cm}|l|}\hline
    
\textbf{Notations}   &  \textbf{Meaning}\\\hline
       $\mathcal{D}_{\mathrm{target}}$ & Target database over which the current NL question is answered\\\hline

       $\mathcal{D}_{\mathrm{train}}$ & Training database having NLQ--gold-SQL examples (used for rule creation)\\\hline
       
        $T_{\!i}.A_{\!j}[k]$ & $k^{\mathit{th}}$ cell value in a table $T_i$ with a column $T_{\!i}$\\\hline

       $\mathit{NLQ}$ \&   $\mathit{gold}_{\mathrm{SQL}}$ & Natural-language question and gold SQL query for $\mathit{NLQ}$\\\hline

       $\mathit{Profile}_{T_{\!i}.A_{\!j}}$ & Profile of column $T_{\!i}.A_{\!j}$\\\hline
       $\mathit{emb}_{T_{\!i}.A_{\!j}}$ & Embedding of $\mathit{Profile}_{T_{\!i}.A_{\!j}}$\\\hline
       $\mathit{Index}_{\mathrm{prof}}$, $\mathit{Index}_{\mathrm{val}}$ & Profile and value vector indices built during preprocessing\\\hline
    
       %$\eta$ & Maximum number of values indexed per text column\\\hline
 
       %$\langle A_i,A_{\!j}\rangle$ & Pair of columns considered for ambiguity\\\hline
       $\mathit{Token}(A)$ & Normalized name tokens of column $A$\\\hline
       $\mathrm{value}(A)$ & Distinct non-\textsf{null} values of column $A$\\\hline
       $\mathit{Pair}_{\mathrm{candidate}}$ & Candidate ambiguous column-pair set\\\hline
       $\mathit{Pair}_{\mathrm{confirm}}$ & Confirmed ambiguous column-pair set\\\hline
      
       $\mathit{Note}_{\langle A_i,A_{\!j}\rangle}$ & Disambiguation note for column pair $\langle A_i,A_{\!j}\rangle$\\\hline

       $\mathcal{N}_{\mathit{NLQ}}$ \& $\mathcal{N}^{\star}_{\mathit{NLQ}}$ & Candidate note and selected relevant note sets for $\mathit{NLQ}$\\\hline
       
       $\mathit{fail}_{\mathrm{candidate}}$, $\mathit{fail}_{\mathrm{final}}$ & Candidate failure explanations and final database-agnostic failures\\\hline
       $\mathit{ErrorGroup}$ & Set of dominant groups of similar failures\\\hline
       $\mathit{Rule}$ & Complete set of synthesized correction rules\\\hline
       $\mathit{rule}_i$ & An individual correction rule in $\mathit{Rule}$\\\hline
       $\mathit{Focused}_{\mathit{NLQ}}$ & Final focused schema for $\mathit{NLQ}$\\\hline
       
       $\mathit{candidate}_{\mathrm{SQL}}$ & A generated candidate SQL\\\hline

       $\mathit{revise}_{\mathrm{SQL}}$ & Candidate SQL after LLM-based revision\\\hline
       $\mathit{correct}_{\mathrm{SQL}}$ & Set of SQL candidates after rule-guided correction\\\hline
       $\mathit{Cluster}_i$, $s_i$ & Execution-result cluster and its representative SQL candidate\\\hline
       % $\mathit{Conf}(s_i)$ & Fraction of corrected candidates in $\mathit{Cluster}_i$\\\hline
       % $K$ & Number of top clusters considered during LLM adjudication\\\hline
       $\mathit{final}_{\mathrm{SQL}}$ & Final SQL query returned by \algname{}\\\hline
       % $\mathrm{EX}$, $\mathrm{UB\text{-}EX}$ & Execution accuracy and upper-bound execution accuracy\\\hline
    \end{tabular}
    \caption{Notations used in the paper.}
    \label{tab:Notations used in the paper}
    \BBB\BBB\BBB
  %  \vspace{-1cm}
\end{table}

\begin{figure*}[t]
%\BBB%\BBB\BB
    \centering
    \includegraphics[scale=0.5]{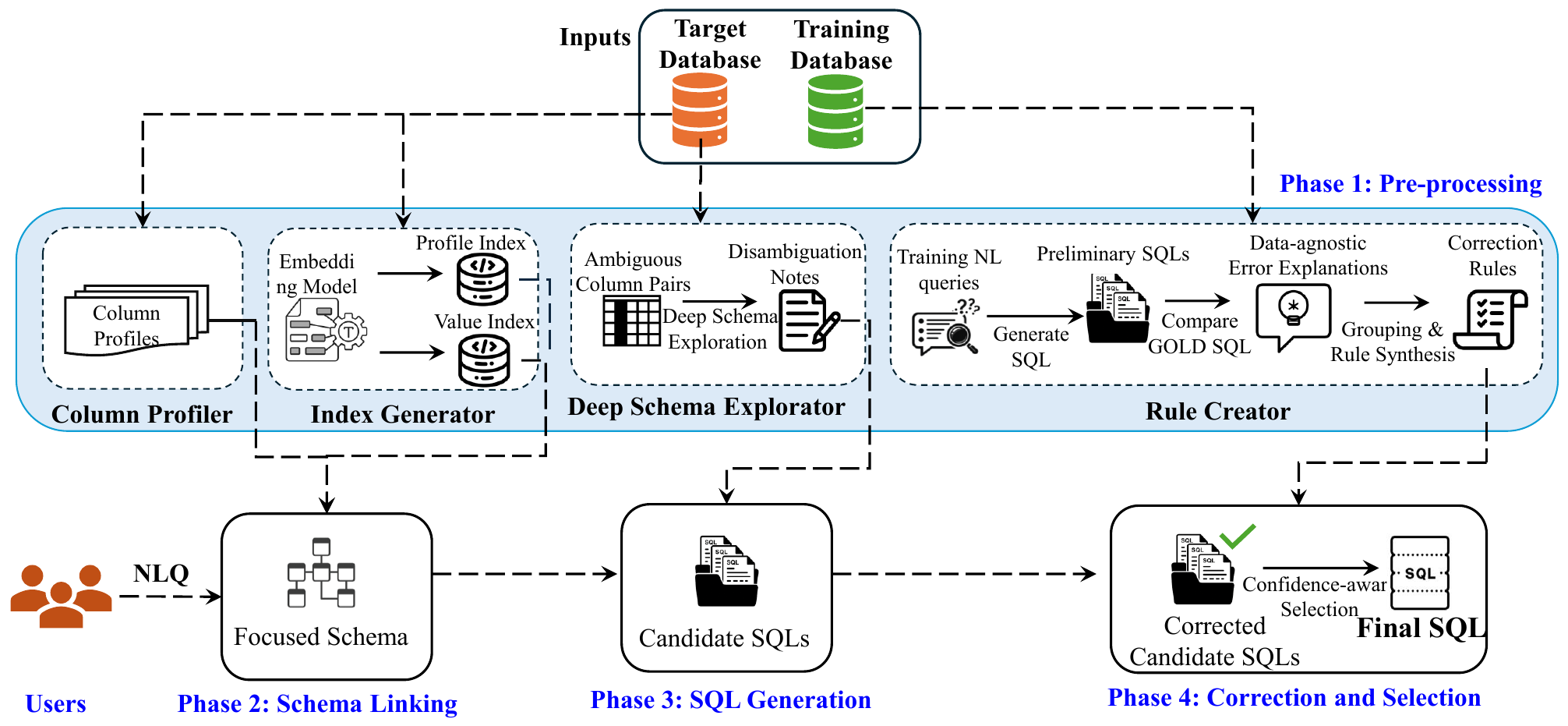}
    %\BBB
    \caption{Overview of \algname{}. 
    %'s offline pre-processing and online use. 
  {\scriptsize\textnormal{Column Profiler, Index Generator, and Deep Schema Explorator process target databases to produce profiles, indices, and disambiguation notes, while Rule Creator mines training databases for correction rules. These artifacts are reused during online Schema Linking, SQL Generation, Correction, and Selection without updating model parameters. }}}
    \label{fig:phase1}
    %\BBB\B
\end{figure*}

\medskip
\noindent
\textbf{Challenge 3: Effective multi-path SQL generation.}
The third challenge is preserving every required condition and relationship when decomposing a complex NL question for SQL generation. A free-form LLM decomposition may omit, hallucinate, or misplace a question element, producing executable SQL that answers an incomplete or different version of the question.

For example, ``\emph{How many examinations for patient \textsf{ID}~1 were diagnosed as \textsf{PSS}?}'' contains two required filters: \textsf{ID = 1} and \textsf{Diagnosis = 'PSS'}. An LLM-generated decomposition may retain the patient filter but drop the diagnosis filter, producing a query equivalent to \textsf{COUNT(*) WHERE ID = 1}. Using {\color{blue}Table~\ref{tab:running-examination}}, this query returns $2$ instead of the correct answer $1$ because it counts both examination records for patient~1.

Prior approaches already use decomposition and multi-path reasoning to diversify SQL generation~\cite{pourreza2023din,pourreza2024chase,li2026deepeye}. However, when a path relies on free-form LLM decomposition, its intermediate steps can still omit or hallucinate a required question element.

% {\color{red}{\bf From this discussion, it was not clear to why ours is novel compared to existing ones. Please add a clear novelty statement compared to previous work. }}

\medskip
\noindent
\textbf{Our solution: Multi-path SQL generation ({\color{blue}\S\ref{sec:sql-generator}}).} 
\algname{} introduces a \emph{\textbf{dependency-tree-based SQL}} intermediate representation. % that decomposes the question deterministically from its grammatical structure rather than relying on an LLM to freely decide the decomposition. 
{%\color{cyan} -- MURAT 
Unlike prior approaches that rely on an LLM to generate the decomposition itself, \algname{} deterministically derives the decomposition from the question's dependency structure and converts it into an SQL-oriented intermediate representation.} 
The obtained dependency tree preserves question's mentions, literals, and their relationships in an SQL skeleton before outputting the final SQL.

To further increase candidate diversity, \algname{} combines this new dependency-tree-based method with two established generation methods: few-shot in-context learning, which uses structurally similar training examples, and divide-and-conquer generation, which solves complex questions by breaking them into simpler sub-questions. These three paths provide complementary ways to generate SQL, making the overall process more effective when one reasoning strategy fails.

For the example above, the dependency tree separately identifies the phrases specifying patient \textsf{ID}=1 and diagnosis \textsf{PSS}. It maps them to the two required filters, \textsf{ID=1} and \textsf{Diagnosis=`PSS'}, in the intermediate SQL representation. The final SQL is then generated from this representation, reducing the risk that either filter is omitted.

\subsection{High-Level Overview of \algname{}}
\label{sec:overview}

\algname{} is divided into four phases, see {\color{blue}Figure~\ref{fig:phase1}}. This section provides an outline of four phases, which are then explained in detail in~{\color{blue}\S\ref{sec:details_of_algo}}.
~We first define the terminology used in this paper.

We refer to the reference SQL query provided by a benchmark together with its corresponding natural language (NL) question as the \emph{\textbf{gold SQL query}} (denoted by $\mathit{gold}_{\mathrm{SQL}}$).
~
We refer to the database on which future NL questions are answered as the \emph{\textbf{target database}} (or \emph{\textbf{test database}}), denoted by $\mathcal{D}_{\mathrm{target}}$. 
\algname{} may inspect only the target database's schema, stored data, and developer-provided documentation. It does \emph{NOT} access or use the target database's gold SQL queries.
~
We refer to a database containing NL questions paired with their corresponding gold SQL queries as a \emph{\textbf{training database}}, $\mathcal{D}_{\mathrm{train}}$. 
\textbf{Training databases and testing databases are disjoint}. Similar to other in-context Text-to-SQL systems (\textit{e}.\textit{g}.,~\cite{li2026deepeye, xie2025opensearch, gao2023text}), \algname{} does not use the training database to update the LLM's parameters or perform any fine-tuning. %Instead, the training database is used only during offline preparation.

\algname{} separates execution into an \emph{\textbf{offline preparation}} stage and an \emph{\textbf{online inference}} stage.
~
\emph{{Offline preparation}} is performed before any NL question is received. It operates on both the training and target databases to construct the information required during inference.
~
\emph{{Online inference}} (or \emph{{SQL generation}}) begins after an NL question over the target database is received. \algname{} generates the corresponding SQL query, which is then compared with the benchmark's gold SQL query to evaluate the system's accuracy.
~
Overall, \textsc{Phase~1} is executed once as an offline preprocessing step, whereas \textsc{Phases~2--4} are executed online for each incoming NL question over the target database.

%A \emph{target database}, also referred to as a test database, is a database 

%A separate \emph{training split} contains databases with NL questions paired with gold SQL and is used only to mine reusable correction rules. No phase updates the LLM's parameters or performs fine-tuning.

\smallskip
\noindent
\textbf{\textsc{Phase~1}: Preprocessing -- An offline phase ({\color{blue}\S\ref{sec:preprocessing}}).}
This phase runs \emph{offline, only once}, before any NL question is asked. 
Inputs to this phase are $\mathcal{D}_{\mathrm{target}}$ and $\mathcal{D}_{\mathrm{train}}$, and the outputs are the per-column profiles, vector indices, disambiguation notes, and correction rules, which are reused throughout SQL query generation.

\textsc{Phase~1} has four components:
(\textit{i})~\emph{Column profiler} that analyzes  contents of $\mathcal{D}_{\mathrm{target}}$ to build a per-column statistics profile;
(\textit{ii})~\emph{Index generator} that builds searchable vector indices over the column profiles and column values of $\mathcal{D}_{\mathrm{target}}$, to help LLM in retrieving semantically relevant columns and values during inference;
(\textit{iii})~\emph{Deep schema explorator} that identifies groups of ambiguous columns and probes $\mathcal{D}_{\mathrm{target}}$ to produce compact disambiguation notes so that confusable columns can be told apart during generation; and
(\textit{iv})~\emph{Rule creator} that mines the $\mathcal{D}_{\mathrm{train}}$ for recurring, \emph{\textbf{database-agnostic}} mismatches between generated and gold SQL and distills them into generalized correction rules to capture recurring LLM failures.

\smallskip
\noindent
\textbf{\textsc{Phase~2}: Schema Linking -- An online phase ({\color{blue}\S\ref{sec:schema-linking}}).}
This phase runs for each NL question. The input to this phase is the NL question together with the full database schema of $\mathcal{D}_{\mathrm{target}}$ and the profiles and vector indices produced by \textsc{Phase~1}. The output is a \emph{focused schema} that contains only the tables and columns relevant to the NL question. 
\textsc{Phase~2} consist of a single component: \emph{schema linker} that takes the input and produces the output, as mentioned before. %that uses the profiles and vector indices (produced by \textsc{Phase~1}) to narrow the full schema to the tables and columns relevant to the NL question, producing a focused schema for more effective SQL generation.

\smallskip
\noindent
\textbf{\textsc{Phase~3}: SQL Generation -- An online phase ({\color{blue}\S\ref{sec:sql-generation}}).}
This phase runs for each NL question. The input to this phase is the NL question, the focused schema from \textsc{Phase~2}, and the disambiguation notes from \textsc{Phase~1}. The output is a set of candidate SQL queries. \textsc{Phase~3} has two components:
(\textit{i})~\emph{Note incorporator} that inserts the only relevant disambiguation notes created by the deep schema explorator in \textsc{Phase~1} into the prompt for SQL generation; and %; since providing all of the notes would be noisy and unhelpful, \algname{} uses an LLM-based gating system to select only the relevant ones; and
(\textit{ii})~\emph{SQL generator} that produces candidate SQLs based on the focused schema produced by \textsc{Phase~2} and the selected disambiguation notes, generating candidates with 
three reasoning strategies: dependency-tree-based intermediate-representation generation, few-shot in-context-learning generation, and divide-and-conquer generation.

\smallskip
\noindent
\textbf{\textsc{Phase~4}: SQL Correction and Selection -- An online phase ({\color{blue}\S\ref{sec:sql-correction-selection}}).}
This phase runs for each NL question. The input to this phase is the set of candidate SQL queries produced by \textsc{Phase~3}. The output is a single final SQL query. 
This phase has two components:
(\textit{i})~\emph{Correction} that repairs generated candidates using deterministic checks, execution feedback, and the database-agnostic correction rules produced in \textsc{Phase~1}; and
(\textit{ii})~\emph{Selection} that chooses the most reliable SQL from the corrected candidate pool.

% \begin{figure*}[t]
%     \centering
%     \includegraphics[width=\textwidth]{Phase1.png}
%     \caption{Phase 1: Pre-processing}
%     \label{fig:phase1}
% \end{figure*}

%======================================================================

\section{Details of \textsc{DexterSQL}}
\label{sec:details_of_algo}

This section explains each phase of \algname{} and its components.  

\subsection{\textsc{Phase~1}: Pre-processing}
\label{sec:preprocessing}

\textsc{Phase~1} consists of four components (see {\color{blue}Figure~\ref{fig:phase1}}):
Column Profiler ({\color{blue}\S\ref{sec:profiling}}), Index Generator ({\color{blue}\S\ref{sec:indexing}}), 
Deep Schema Explorator ({\color{blue}\S\ref{sec:deep-exploration}}), and Rule Creator ({\color{blue}\S\ref{sec:rule-creation}}); as discussed in {\color{blue}\S\ref{sec:overview}}.  
Below, we explain each component.

\subsubsection{\textbf{Column Profiler}}
\label{sec:profiling}

\textbf{\emph{Purpose.}} For a given NL question, during SQL generation based only on the database schema, an LLM cannot understand the underlying semantics of the database for three reasons: 
(\textit{i})~the schema carries limited information about the data,  
(\textit{ii})~column identifiers are often abbreviated or ambiguous and declared datatypes (\textit{e}.\textit{g}., \texttt{TEXT}) reveal little about the actual format or domain of the stored values,  and 
(\textit{iii})~a database may contain columns with similar names or related meanings, making it harder to identify the correct columns during schema understanding and SQL generation. 
To solve these problems, \algname{} provides the LLM with fine-grained, data-derived information about each column through column profiler. %, which analyzes the database contents during preprocessing.

%summarizes what the column represents and what its values look like, giving the LLM the context needed to map natural-language queries to the appropriate schema elements.

\noindent\textbf{\emph{Working of column profiler.}}
For each column $A_{\!j}$ of each table $T_{\!i}$ (\textit{i}.\textit{e}., $T_{\!i}.A_{\!j}$) of the \emph{\textbf{target database}}, the column profiler builds a profile, $\mathit{Profile}_{T_{\!i}.A_{\!j}}$,  from the column's data, containing: % two kinds of information:
\begin{enumerate}
[leftmargin=0.01in, noitemsep,nolistsep]
    \item \textbf{Cardinality statistics:} include the row count, the number of \textsf{null} and non-\textsf{null} values, the number of distinct values, the range of values, and the minimum/ maximum values. These indicate whether a column is sparse (many \textsf{null}s), unique or key-like (distinct count close to the row count), categorical (few distinct values), or numeric within a bounded range.

    \item \textbf{Representative values:} a sample of the column's values, together with its most frequent values and their counts. These give LLM concrete examples of what the column actually stores and how its values are formatted. %, which its name and declared type alone do not convey.
\end{enumerate}

When available, $\mathit{Profile}_{T_{\!i}.A_{\!j}}$ also contains developer-provided column descriptions, value notes, and any other %content-derived statistics with 
human provided documentation in a single profile.

\subsubsection{\textbf{Index Generator}}
\label{sec:indexing}

\textbf{\emph{Purpose.}}
To map an NL query to columns, \algname{} needs to quickly 
(\textit{i})~retrieve semantically relevant columns whose profiles match the NL question, and 
(\textit{ii})~locate columns that contain specific names, categories, dates, or numeric values from the NL question. 
To do so, \algname{} builds two complementary indices over the  \textbf{\emph{target database}}---a profile index $\mathit{Index}_{\mathrm{prof}}$ for semantic retrieval over column profiles and a value index $\mathit{Index}_{\mathrm{val}}$ for literal-value retrieval over column values.

\noindent
\textbf{\emph{Working of index generator.}} %To this end, \algname{} builds two complementary indices---a profile index $\mathit{Index}_{\mathrm{prof}}$ for semantic retrieval over column profiles and a value index $\mathit{Index}_{\mathrm{val}}$ for literal-value retrieval over column values. \algname{} constructs the two indices as follows:
The two indices are created as follows:
\begin{enumerate}[leftmargin=0.01in, noitemsep,nolistsep]
    \item \textbf{Profile index.}
    Index generator embeds, for each column $T_{\!i}.A_{\!j}$,  its profile $\mathit{Profile}_{T_{\!i}.A_{\!j}}$ into a dense vector $\mathit{emb}_{T_{\!i}.A_{\!j}}=\mathrm{Embed}(\mathit{Profile}_{T_{\!i}.A_{\!j}})$ using an embedding model and stores it in a FAISS vector index~\cite{douze2024faiss}, resulting in profile index:  $\mathit{Index}_{\mathrm{prof}}=\{\mathit{emb}_{T_{\!i}.A_{\!j}} : T_{\!i}.A_{\!j} \in \mathcal{D}_{\mathrm{target}}\}$.
    % Given an NL query, the query is embedded into the same vector space, and the top-$k$ columns with the most similar profile vectors are retrieved. This surfaces relevant columns even when the query does not mention the column name directly.

    \item \textbf{Value vector index.}
    This step 
   builds  cosine-similarity vector index~\cite{reimers2019sentence} on distinct non-\textsf{null} values in each \emph{text-typed} column.
  Text values are indexed because semantic matching can connect an NL phrase to a differently worded database value, whereas numeric values are handled via exact values and ranges. 
    To bound the cost, at most $\eta$ (10,000 in our system) values per column are embedded, normalized, and stored in $\mathit{Index}_{\mathrm{val}}$.
    % At inference, literals extracted from the query are embedded and used to retrieve the top-$k$ closest values, which are added back as value examples to help identify the correct column and filter literal.
\end{enumerate}

Together, the profiles $\mathit{Profile}_{T_{\!i}.A_{\!j}}$ and the indices $\mathit{Index}_{\mathrm{prof}}$ and $\mathit{Index}_{\mathrm{val}}$ form the precomputed artifacts that \algname{} uses during \textsc{Phase}~2 schema linking ({\color{blue}\S\ref{sec:schema-linking}}).

\subsubsection{\textbf{Deep Schema Explorator}}
\label{sec:deep-exploration}
\emph{\textbf{Purpose.}} A key limitation of existing LLM-based Text-to-SQL systems during SQL query generation for a given NL query is the lack of exploratory information about how columns relate to one another to the underlying LLM %, even if provide a sample of column-values to LLM 
(as we discussed in Challenge~1; see {\color{blue}\S\ref{subsection:challenges_and_our_solutions}}). 

A trivial solution is to copy all column values into the LLM context. However, it is neither practical nor helpful (specifically for ambiguous columns) and introduces noise.
To address this, \algname{} \emph{identifies only column pairs that may be ambiguous} and. For each such pair $\langle A_i,A_{\!j}\rangle$, \algname{}  produces a disambiguation note $\mathit{Note}_{\langle A_i,A_{\!j}\rangle}$ (as the output of this component) that characterizes how the two columns differ and when each should be used, (supplied as context for relevant questions during SQL generation). 

\noindent
\emph{\textbf{Working of deep schema explorator.}}
To uncover unclear relationships between columns, Deep Schema Explorator follows a four-step process executed over the \emph{\textbf{target database}}. 
\textsc{Step}~1 detects candidate column pairs that may be ambiguous to an LLM and outputs {\color{black}$\mathit{Pair}_{\mathrm{candidate}}$}. 
\textsc{Step}~2 filters these candidates using LLM triage and outputs the confirmed ambiguous set {\color{black}$\mathit{Pair}_{\mathrm{confirm}}$}. 
\textsc{Step}~3 investigates each confirmed pair by determining %comparing its 
individual column-level and joint column-level statistics and producing evidence on the relationship of confirmed ambiguous column pairs.
%, {\color{black}including sparsity (\textit{i}.\textit{e}., how many nulls or missing values in a column)} and frequent values; when a join path exists between the corresponding tables, it also compares the values that co-occur after joining the tables and measures how many joined rows appear per entity. 
\textsc{Step}~4 summarizes evidence from \textsc{Step}~3 into an instruction-oriented disambiguation note $\mathit{Note}_{\langle A_i,A_{\!j}\rangle}$ that instructs SQL-generating LLM how the columns differ and when each one should be used. Below, we provide details of each step.

\smallskip
\noindent\textbf{\textsc{Step~1: Candidate generation.}}
\emph{Purpose.} Since deep investigation over all possible column pairs is infeasible for large databases, we need %candidate generation uses 
cheap deterministic signals to identify only plausible ambiguous pairs.

\emph{Method.} The input to this step is the target database schema  together with the column profiles $\mathit{Profile}_{T_{\!i}.A_i}$ produced by the column profiler 
(\textcolor{blue}{\S\ref{sec:profiling}}), and the output is the candidate pair set $\mathit{Pair}_{\mathrm{candidate}}$, which is provided to \textsc{Step~2}. 
~
%\algname{}  
$\mathit{Pair}_{\mathrm{candidate}}$ is constructed using two deterministic signals (rather than exhaustively investigating every column pair):

\begin{enumerate}[leftmargin=0.01in, noitemsep,nolistsep]
    \item 
\textbf{Checking \emph{name similarity}}: a pair $\langle A_i,A_{\!j}\rangle$ is added if the column names match or if their normalized name-token sets $\mathit{Token}(A_i)$ and $\mathit{Token}(A_{\!j})$ have high overlap, where $\mathit{Token}(A)$ denotes the set of normalized tokens~\cite{rai2020study} obtained from column name $A$. 

\item \textbf{Checking \emph{profile similarity}}: using  $\mathit{Profile}_{T_{\!i}.A_i}$ and $\mathit{Profile}_{T_j.A_{\!j}}$ from {\color{blue}\S\ref{sec:profiling}}, now, we % step %\algname{}
embed each profile and add the pair when the cosine similarity between the two profile embeddings exceeds a threshold. Finally, this step removes structurally obvious non-ambiguities, such as declared primary-key/foreign-key pairs and identifier columns. The remaining pairs form $\mathit{Pair}_{\mathrm{candidate}}$.

\end{enumerate}

\emph{Example.} In the running example in \textcolor{blue}{\S\ref{subsection:challenges_and_our_solutions}}, \textsf{Examination.Diagnosis} and \textsf{Patient.Diagnosis} share the same column name, so name similarity adds $\langle\textsf{Examination.Diagnosis},\textsf{Patient.Diagnosis}\rangle$ to $\mathit{Pair}_{\mathrm{candidate}}$. In contrast, obvious identifier pairs such as \textsf{Examination.ID} and \textsf{Patient.ID} are removed because their relationship is already explained by the primary-key/foreign-key structure.

\smallskip
\noindent\textbf{\textsc{Step~2: LLM triage.}}
\emph{Purpose.} 
% For instance, a \textsf{Name} column appearing in both a \textsf{Patient} table and a \textsf{Doctor} table would be flagged because the names match exactly, but an LLM would realistically understand which column a question intends, so the pair is not genuinely ambiguous.
The candidate set $\mathit{Pair}_{\mathrm{candidate}}$ may contain column pairs that look similar by name or profile but are not actually confusing to LLM for SQL generation. LLM triage filters these candidates and keeps only the genuinely ambiguous pairs.

\emph{Method.} The inputs to this step are the candidate pair set $\mathit{Pair}_{\mathrm{candidate}}$ from \textsc{Step~1}, the column profiles produced by Column Profiler ({\color{blue}\S\ref{sec:profiling}}), the database schema, including primary-key/foreign-key relationships, and any additional database context, \textit{e}.\textit{g}., subject-matter-expert metadata supplied with the benchmark. 
The output is the confirmed ambiguous set $\mathit{Pair}_{\mathrm{confirm}}\subseteq\mathit{Pair}_{\mathrm{candidate}}$, which is passed to \textsc{Step~3} for deep investigation.

For each pair $\langle A_i,A_{\!j}\rangle\in\mathit{Pair}_{\mathrm{candidate}}$, \algname{} uses an LLM-as-judge to decide whether a realistic NL question could plausibly refer to either column and lead the SQL-generating LLM to pick the wrong one. To make the judgment effective, \algname{} evaluates each pair under three prompts with increasing levels of information: (\textit{i})~the column profiles and sample values produced by Column Profiler; (\textit{ii})~the profiles with relevant primary-key/foreign-key information; and (\textit{iii})~the full schema, profiles, and additional database context. Each prompt returns a binary verdict, and \algname{} takes a majority vote.

\emph{Example.} For $\langle\textsf{Examination.Diagnosis},\textsf{Patient.Diagnosis}\rangle$, \algname{} asks the LLM judge the same ambiguity question under three prompt variants. In the first prompt, it shows only the two column profiles and sample values, such as \textsf{SLE}, and asks whether an NL question mentioning ``diagnosis'' could plausibly refer to either column. In the second prompt, it adds relevant primary-key/foreign-key context, so the judge can see that \textsf{Examination.Diagnosis} belongs to examination rows while \textsf{Patient.Diagnosis} belongs to patient rows, and asks whether this distinction is still easy for a SQL-generating LLM to confuse. In the third prompt, it shows the full schema with profiles and context and asks for the same binary ambiguity verdict in the complete database setting. Since the prompt variants agree that the pair can confuse SQL generation, the majority vote keeps it in $\mathit{Pair}_{\mathrm{confirm}}$.

\smallskip
\noindent\textbf{\textsc{Step~3: Deep investigation.}}
\emph{Purpose.} Given the final set of ambiguous pairs $\mathit{Pair}_{\mathrm{confirm}}$, this step aims to understand the underlying relationship between each pair of columns --- how the two columns' values are distributed
and how their distributions vary across the joined rows if there is a join path between their tables.

\emph{Method.} The input to this step is the confirmed ambiguous set $\mathit{Pair}_{\mathrm{confirm}}$ from \textsc{Step}~2. 
For each pair of columns $\langle A_i,A_{\!j}\rangle \in \mathit{Pair}_{\mathrm{confirm}}$, the output is evidence describing their individual distributions and, when applicable, their distribution across a join path. These per-pair column evidence are provided to \textsc{Step~4}. 

%Since the objective of this step is 
To determine a relationship between each confirmed column pair ${\langle A_i,A_{\!j}\rangle\in\mathit{Pair}_{\mathrm{confirm}}}$, where $A_i$ belongs to table $T_{\!i}$ and $A_{\!j}$ belongs to table $T_j$, we %perform the following: 
find 
%we determine the relationship in two ways: 
(\textit{i})~the individual column data distributions and 
(\textit{ii})~ the join-conditioned data distribution across the columns when their tables can be connected through a join path, as follows:

%These two explain whether the two columns represent the same concept at different related granularities,  but distinct concepts, or unrelated values that merely look similar.
%How each view is computed is described below.

\begin{enumerate}[leftmargin=0.01in, noitemsep,nolistsep]
    \item 
  \textbf{\emph{Determining data distribution within a column.}}
 For each column $A_\ast$ in a confirmed ambiguous pair, where $\ast\in\{i,j\}$, this step takes all statistics produced by % the the following measurements already computed by 
Column Profiler ({\color{blue}\S\ref{sec:profiling}}). % $\mathrm{rows}(A_\ast)$ for the row count, $\mathrm{value}(A_\ast)$ for distinct non-\textsf{null} values, $\mathrm{nulls}(A_\ast)$ for missing values, $\mathrm{range}(A_\ast)$ for the value range, and $\mathrm{freq}(A_\ast)$ for frequent values.
Furthermore, for each confirmed ambiguous pair $\{A_i,A_{\!j}\}$, this step computes the distinct non-\textsf{null} value sets $\mathrm{value}(A_i)$ and $\mathrm{value}(A_{\!j})$ and their Jaccard similarity:
$$
  \mathrm{overlap}(A_i,A_{\!j})=
  \frac{|\mathrm{value}(A_i)\cap \mathrm{value}(A_{\!j})|}
       {|\mathrm{value}(A_i)\cup \mathrm{value}(A_{\!j})|}.
$$
For text columns, this step also computes a case-insensitive version of the overlap. These measurements show whether the columns use similar value domains despite their potentially different roles.

\item
 \textbf{\emph{Determining data distribution across columns.}} 
This analysis is necessary because two columns may have similar names and value distributions, yet their semantic roles often become apparent only after examining how their values interact through PK/FK relationships. We use $T_{\!i}\bowtie_{T_{\!i}.X=T_j.X}T_j$ to denote a join between $T_{\!i}$ and $T_j$ based on a primary-key/foreign-key (PK/FK) relationship.

{\color{black}
To capture the relational characteristics of ambiguous columns, we compute three join-path measurements. These measurements quantify 
(\textit{i})~how many rows from one table participate in the join, (\textit{ii})~the average multiplicity of the join relationship, and (\textit{iii})~how frequently the ambiguous column values agree after the join. Together, these signals help distinguish columns that appear similar when viewed independently but serve different semantic roles within the database schema.
}

We assume that $T_{\!i}.X$ is the primary key and $T_j.X$ is the corresponding foreign key. {\color{black}The same measurements are also computed when $T_{\!i}$ and $T_j$ are connected through one or more intermediate tables along a PK/FK join path.}

{\color{black}
\emph{Coverage} measures the fraction of rows from $T_{\!i}$ that participate in at least one join with $T_j$:
}
\[
  \mathrm{coverage}(A_i,A_{\!j})=
  \frac{|\pi_{T_{\!i}.X}(T_{\!i}\bowtie_{T_{\!i}.X=T_j.X}T_j)|}
       {|\pi_{T_{\!i}.X}(T_{\!i})|}.
\]

\emph{Fan-out} measures, on average, how many joined rows are produced for each participating row from $T_{\!i}$. {\color{black}This captures the cardinality of the relationship and indicates whether a matched row in $T_{\!i}$ typically corresponds to a single row or multiple rows in $T_j$:}
\[
  \mathrm{fanout}(A_i,A_{\!j})=
  \frac{|T_{\!i}\bowtie_{T_{\!i}.X=T_j.X}T_j|}
       {|\pi_{T_{\!i}.X}(T_{\!i}\bowtie_{T_{\!i}.X=T_j.X}T_j)|}.
\]

\emph{Agreement} measures how frequently the two ambiguous columns contain identical values among the joined tuples. A high agreement score suggests that the columns may represent the same underlying concept despite appearing in different tables:
\[
  \mathrm{agreement}(A_i,A_{\!j})=
  \frac{|\sigma_{T_{\!i}.A_i=T_j.A_{\!j}}(T_{\!i}\bowtie_{T_{\!i}.X=T_j.X}T_j)|}
       {|T_{\!i}\bowtie_{T_{\!i}.X=T_j.X}T_j|}.
\]

{\color{black}
If no PK/FK join path exists between $T_{\!i}$ and $T_j$, \algname{} omits these join-path measurements and relies solely on the column-level distributional statistics described earlier.
}

\end{enumerate}

\emph{Example.} For the diagnosis pair, let $A_i=\textsf{Patient.Diagnosis}$ and $A_{\!j}=\textsf{Examination.Diagnosis}$. The column-level probe first compares their distributions, including nulls, frequent values, and value overlap, and finds that both columns share diagnosis labels such as \textsf{SLE}, \textsf{PSS}, and \textsf{APS}.

The join-path probe then finds a direct path \textsf{Examination.ID}$\rightarrow$\textsf{Patient.ID}. After joining through this path, it computes $\mathrm{coverage}(A_i,A_{\!j})$, %=763/1238=61.6\%$, 
$\mathrm{fanout}(A_i,A_{\!j})$, %$=1.1$, 
and $\mathrm{agreement}(A_i,A_{\!j})$%$=32.3\%$ 
over the overlapping joined rows. These measurements show that the two columns share a diagnosis vocabulary, but \textsf{Examination.\allowbreak Diagnosis} covers only a subset of patients and often disagrees with \textsf{Patient.\allowbreak Diagnosis} after joining. This evidence is passed to note synthesis, which summarizes that \textsf{Patient.\allowbreak Diagnosis} should be used for patient-level final-diagnosis questions, while \textsf{Examination.\allowbreak Diagnosis} should be used for examination-level diagnosis questions.

\smallskip
\noindent\textbf{\textsc{Step~4: Note synthesis.}}
\emph{Purpose.} TPrevious step produces useful but low-level evidence (not fine-grained enough to place directly in SQL generation prompt). 
This step converts the output of \textsc{Step}~3 into concise guidance that LLM uses to disambiguate columns.

\emph{Method.} Input to this step is the evidence computed in \textsc{Step~3} for each confirmed pair $\langle A_i,A_{\!j}\rangle$; 
%, including the column-level distributional statistics, $\mathrm{overlap}(A_i,A_{\!j})$, and, when a join path exists, $\mathrm{coverage}(A_i,A_{\!j})$, $\mathrm{fanout}(A_i,A_{\!j})$, and $\mathrm{agreement}(A_i,A_{\!j})$. 
and the output is a disambiguation note $\mathit{Note}_{\langle A_i,A_{\!j}\rangle}$ for each ambiguous pair. 

For each pair $\langle A_i,A_{\!j}\rangle\in\mathit{Pair}_{\mathrm{confirm}}$, this step uses an LLM to summarize the evidence obtained in \textsc{Step}~3 into guidance $\mathit{Note}_{\langle A_i,A_{\!j}\rangle}$  that states how the columns differ, when each column should be used, and any other information needed to help the LLM make the correct decision during inference. 
If multiple ambiguous pairs share columns, this step groups them into a \emph{\textbf{confusable family}} and creates one \emph{family-level note}. The final output is a per-database noteset $\mathcal{N}(\mathcal{D}_{\mathrm{target}})$, computed \emph{\textbf{once, offline}} and reused during inference as discussed in the schema-linking phase ({\color{blue}\S\ref{sec:schema-linking}}).

%{\color{red} now it is confusing ---- u say a per-database noteset----however, this is offline and future query's schema agnostic....so why i create per db note that for all training dataset, because all such dataset are useless for ur any future queries....u need only collection of }

\emph{Example.} For the diagnosis pair, note synthesis looks at the Step~3 evidence, including value overlap, coverage, fan-out, and agreement, and infers that the two columns share a diagnosis vocabulary but play different roles: \textsf{Patient.Diagnosis} is a patient-level final diagnosis, while \textsf{Examination.Diagnosis} is an examination-level diagnosis tied to a specific examination event. To make this distinction clear during inference, the resulting $\mathit{Note}_{\langle A_i,A_{\!j}\rangle}$ describes the role of each column and gives usage guidance: use \textsf{Patient.Diagnosis} when the question asks for a patient's diagnosis or final diagnosis, and use \textsf{Examination.Diagnosis} when the question asks about diagnosis information associated with an examination.

\subsubsection{\textbf{Rule Creator}}
\label{sec:rule-creation}

\emph{\textbf{Purpose.}} To overcome recurring SQL-generation errors of the underlying LLM, \algname{} creates rules using only training databases, as discussed in Challenge~2 ({\color{blue}\S\ref{subsection:challenges_and_our_solutions}}). These rules are database-agnostic and created only once, offline, without fine-tuning the model.

\noindent
\emph{\textbf{Working of rule creator.}}
Rule Creator follows a four-step process. Step~1 samples a subset of training questions, mines generated SQLs that disagree with the gold SQL, and explains the observed failures. Step~2 filters these explanations to keep reusable database-agnostic failures. Step~3 clusters similar failures into dominant error groups. Step~4 converts each dominant group into a correction rule $\mathit{rule}_i$. The output is the rule set $\mathit{Rule}$, where each rule provides a directive indicating when it applies and how to fix the corresponding SQL error. We describe each step below.

\smallskip
\noindent\textbf{\textsc{Step~1: Error mining.}}
\emph{Purpose.} The objective of this step is to find concrete LLM failures on the training databases, which provide the raw material for rule creation.

\emph{Method.} The inputs are training examples $\langle\mathcal{D}_{\mathrm{train}},\mathit{NLQ},\mathit{gold}_{\mathrm{SQL}}\rangle$ and an underlying LLM. 
The output is the candidate failure-explanation set $\mathit{fail}_{\mathrm{candidate}}$.

This step samples questions associated with $\mathcal{D}_{\mathrm{train}}$ using difficulty labels when available. If such labels are unavailable, this step estimates difficulty based on the gold SQL structure, such as whether the query contains joins, nesting, aggregation, grouping, ordering, or set operations, so that the sampled questions cover a range of SQL complexities.

Then, for each sampled question $\mathit{NLQ}$, this step provides the question, the schema context from $\mathcal{D}_{\mathrm{train}}$, and any question-specific evidence to  LLM, and asks it to generate multiple candidate SQLs.
Each generated $\mathit{candidate}_{\mathrm{SQL}}$ is executed on $\mathcal{D}_{\mathrm{train}}$ and compared with $\mathit{gold}_{\mathrm{SQL}}$. A candidate whose execution result matches $\mathit{gold}_{\mathrm{SQL}}$ is discarded because it answers the NL question correctly.

For each remaining incorrect candidate, \algname{} uses an LLM to compare $\mathit{candidate}_{\mathrm{SQL}}$ with $\mathit{gold}_{\mathrm{SQL}}$ under the relevant context and explain why the generated SQL fails. If one candidate contains multiple independent problems, \algname{} keeps them as separate per-question failure explanations. Repeating this process for all sampled questions produces $\mathit{fail}_{\mathrm{candidate}}$.

\emph{Example.} For the Challenge~2 ratio question in the running example ({\color{blue}\S\ref{subsection:challenges_and_our_solutions}}), $\mathit{gold}_{\mathrm{SQL}}$ casts one side of the division to \textsf{REAL} and returns $0.5$. The LLM may instead generate $\mathit{candidate}_{\mathrm{SQL}}$ as \textsf{SUM(Diagnosis = 'PSS') / SUM(Diagnosis = 'SLE')} over \textsf{Examination}. Although it selects the correct table and filters, this candidate returns $0$ because both operands are integer-valued. Its execution result therefore differs from that of $\mathit{gold}_{\mathrm{SQL}}$, so \algname{} retains the candidate and adds an explanation: The query omits a floating-point conversion, leaving both aggregate expressions as integers; SQLite therefore performs integer division and truncates the fractional result to $0$.

\smallskip
\noindent\textbf{\textsc{Step~2: Database-agnostic error isolation.}}
\emph{Purpose.} Not every mismatch should be used to create a corrective rule. Some errors arise from missing schema information, incorrect schema linking, or a database-specific misunderstanding; such errors are tied to a particular database and do not generalize. This step therefore keeps only database-agnostic errors.

\emph{Method.} The input to this step is $\mathit{fail}_{\mathrm{candidate}}$ from \textsc{Step~1}; the output is the final database-agnostic failure set $\mathit{fail}_{\mathrm{final}}\subseteq\mathit{fail}_{\mathrm{candidate}}$. For each explanation, \algname{} uses an LLM to decide whether the failure reflects a reusable SQL-formulation issue rather than a database-specific artifact such as a particular table, column, literal, or schema-linking mistake. Explanations judged to be database-specific are discarded, while database-agnostic explanations are kept in $\mathit{fail}_{\mathrm{final}}$ for clustering.

\emph{Example.} The missing-\textsf{REAL}-cast explanation from \textsc{Step~1} is retained in $\mathit{fail}_{\mathrm{final}}$: the same integer-division error can occur in any database whenever a ratio divides two integer aggregates. In contrast, consider a patient-diagnosis question whose gold SQL uses \textsf{Patient.Diagnosis}, but whose generated SQL uses \textsf{Examination.\allowbreak Diagnosis}. This mismatch arises from the meanings of these particular columns, so \algname{} discards its explanation as a database-specific schema-linking error.

\smallskip
\noindent\textbf{\textsc{Step~3: Clustering errors.}}
\emph{Purpose.} 
We need to cluster similar database-agnostic explanations to form dominant error groups.

\emph{Method.} The input to this step is $\mathit{fail}_{\mathrm{final}}$ from \textsc{Step~2}; the output is $\mathit{ErrorGroup}$, the set of dominant groups in which similar failure explanations are clustered together. \algname{} clusters the explanations hierarchically with an LLM. Within each training database, it divides the corresponding explanations into manageable batches and asks the LLM to group explanations that describe the same underlying SQL-formulation error. It then compares and merges similar batch-level groups to form one set of error groups for that database. \algname{} repeats this comparison-and-merging process across training databases, merging groups that describe the same underlying error. Groups with too few supporting explanations are discarded, and the remaining dominant groups form $\mathit{ErrorGroup}$ for \textsc{Step~4}.

\emph{Example.} The Challenge~2 explanation, which identifies a missing floating-point conversion as the cause of a truncated ratio, is first grouped with similar explanations from \textsf{thrombosis\_prediction}. During cross-database merging, this group is merged with a group from another database containing similar truncated-ratio errors.

\smallskip
\noindent\textbf{\textsc{Step~4: Rule synthesis.}}
\emph{Purpose.} Rule synthesis converts each dominant error group into a reusable optimal correction rule.

\emph{Method.} The input %to this step 
is the dominant error group $\mathit{ErrorGroup}$ from \textsc{Step~3}, and the output is the rule set $\mathit{Rule}$ containing correction rules $\mathit{rule}_i$, which are used during inference to identify LLM SQL failures and correct them. For each dominant error group in $\mathit{ErrorGroup}$, \algname{} uses an LLM to synthesize a canonical rule:
\[
  \mathit{rule}_i=\langle
    \text{gist},\ \text{bad-pattern},\ \text{correct-pattern},\ \text{fix}
  \rangle.
\]
Here, \textit{gist} summarizes the recurring failure and the situation in which it occurs, \textit{bad-pattern} identifies the faulty SQL formulation to detect, \textit{correct-pattern} gives the corresponding correct formulation, and \textit{fix} provides the instruction for transforming the faulty formulation into the correct one. 

{
An LLM-synthesized rule may contain incorrect error reasoning or patterns, hallucinated instructions, or otherwise be suboptimal. Thus, applying such  rules may harm a generated SQL instead of correcting it. To address this issue, this step also validates and refines each rule before adding it to the final rule set. For this, this step samples a mixed batch of generated candidate SQLs from $\mathcal{D}_{\mathrm{train}}$. The batch contains correct candidates whose execution results match $\mathit{gold}_{\mathrm{SQL}}$ and incorrect candidates whose execution results do not match $\mathit{gold}_{\mathrm{SQL}}$. An LLM identifies the candidates to which $\mathit{rule}_i$ applies, using the same rule-relevance selection procedure later employed in \textsc{Step~3} of Correction in \textsc{Phase~4} ({\color{blue}\S\ref{sec:correction}}), and revises each selected candidate using the rule. 
Then, this step executes each rule-corrected SQL and compares its result with $\mathit{gold}_{\mathrm{SQL}}$ and uses two signals to find whether the rule is suitable or requires further refinement. 
A case is \emph{helped} when an originally incorrect candidate becomes correct after applying the rule, and a case is \emph{harmed} when an originally correct candidate becomes incorrect. 
A rule is accepted only if it helps at least one candidate and its harm ratio, $\mathit{harm\_ratio}=|\mathrm{harmed}|/|\mathrm{helped}|$, is less than $0.2$.

If a rule fails validation, this step asks LLM to refine it by providing cases containing $\mathit{NLQ}$, $\mathit{gold}_{\mathrm{SQL}}$, the SQL before and after rule application, and whether the rule helped, harmed, or left the case unchanged. 
LLM reflects on these outcomes and revises the rule's reasoning and correction instruction. The refined rule is validated again for at most three rounds. A rule is added to $\mathit{Rule}$ once it satisfies the threshold; otherwise, discarded.}

\emph{Example.} From the truncated-ratio group, \algname{} creates the following rule. \textit{Gist:} A ratio formed by dividing integer counts or aggregates omits a floating-point conversion and may be truncated. \textit{Bad pattern:} \textsf{SUM(cond\_a) / SUM(cond\_b)}. \textit{Correct pattern:} \textsf{CAST(SUM(cond\_a) AS REAL) / SUM(cond\_b)}. \textit{Fix:} When this pattern is used to compute a ratio, convert one operand to a floating-point value before division. 
% The resulting $\mathit{rule}_i$ is added to $\mathit{Rule}$ for later SQL correction. 
{The rule is added to $\mathit{Rule}$ if it helps at least one validation case and its $\mathit{harm\_ratio}<0.2$; otherwise, it is refined and revalidated for up to three rounds.}

\subsection{\textsc{Phase~2}: Schema Linking}
\label{sec:schema-linking}

\emph{\textbf{Purpose.}} Schema linking maps an NL question to the part of the database schema needed to answer it. This is necessary because real databases may contain many tables and columns, and giving the full schema to the LLM can introduce irrelevant columns, confuse similar schema elements, and make SQL generation less reliable. The goal is therefore to construct a focused schema $\mathit{Focused}_{\mathit{NLQ}}$ that preserves the tables and columns needed for the question while removing unnecessary schema context.

\noindent
\emph{\textbf{Working.}} Given an NL question $\mathit{NLQ}$ over target database $\mathcal{D}_{\mathrm{target}}$, the inputs to this phase are schema of $\mathcal{D}_{\mathrm{target}}$,
the column profiles $\mathit{Profile}_{T_i.A_j}$ produced by Column Profiler ({\color{blue}\S\ref{sec:profiling}}), 
the profile and value indices $\mathit{Index}_{\mathrm{prof}}$ and $\mathit{Index}_{\mathrm{val}}$ produced by Index Generator ({\color{blue}\S\ref{sec:indexing}}),  and any additional schema information, such as subject-matter-expert metadata supplied with the benchmark. The output is a focused schema, $\mathit{Focused}_{\mathit{NLQ}}$, containing the tables and columns needed to answer $\mathit{NLQ}$, which is provided to \textsc{Phase~3}.
\algname{} constructs this output in two steps.

\smallskip
\noindent\textbf{\textsc{Step~1: Preliminary focused-schema construction.}}
\emph{Purpose.} We need to prune several of the tables/columns at the coarse-grain level, since the schema of $\mathcal{D}_{\mathrm{target}}$ may contain many irrelevant columns/tables to NLQ. 
%this step creates a compact preliminary schema that provides \textsc{Step~2} with a question-relevant starting point for bidirectional schema linking.

\emph{Method.} All inputs to this phase are provided to the first step, and 
%step are $\mathit{NLQ}$, the column profiles, and the profile and value indices; 
the output is a \emph{preliminary focused schema} $\mathit{Focused}^{\prime}_{\mathit{NLQ}}$.

This step selects columns by  
%\algname{} uses two complementary retrieval signals.
%First, it 
(\textit{i})~searching $\mathit{Index}_{\mathrm{prof}}$ for columns whose profile descriptions are semantically similar to $\mathit{NLQ}$ --- this retrieves columns that match the meaning of the question even when their names are not explicitly mentioned, and %then, 
%Second, \algname{} 
(\textit{ii})~extracting literals from $\mathit{NLQ}$ and searches $\mathit{Index}_{\mathrm{val}}$ for similar values stored in $\mathcal{D}_{\mathrm{target}}$ ---  value matching % is an actual database value retrieved because it is similar to a question literal; 
helps to select a (candidate) column containing that value.
%\footnote{\scriptsize This process is similar to~\cite{shkapenyuk2025automatic}.}
%,  is therefore a candidate for expressing the corresponding SQL predicate. 

The columns returned by this way %both signals, together 
with their tables form $\mathit{Focused}^{\prime}_{\mathit{NLQ}}$. The retrieved database values are also retained as grounding information for SQL generation.

\smallskip
\noindent\textbf{\textsc{Step~2: Final focused-schema construction.}}
\emph{Purpose.} Profile similarity and literal-value matching in \textsc{Step~1} may miss required schema elements when they are only indirectly implied by $\mathit{NLQ}$ or are needed solely to connect tables. They may also retrieve semantically similar columns that are not required by the intended SQL query. This step therefore uses the preliminary focused schema to guide bidirectional schema linking and determine the final set of tables and columns required to answer $\mathit{NLQ}$.

%to guide {\color{red}no one knows the following term -- write exact meaning ---bidirectional schema linking} and produce the final set of tables and columns required to answer $\mathit{NLQ}$.

\emph{Method.} The inputs are $\mathit{NLQ}$ and the full schema of $\mathcal{D}_{\mathrm{target}}$ supplied to this phase; $\mathit{Focused}^{\prime}_{\mathit{NLQ}}$ produced in \textsc{Step~1}; the column profiles $\mathit{Profile}_{T_i.A_j}$ produced by Column Profiler ({\color{blue}\S\ref{sec:profiling}}); the value index $\mathit{Index}_{\mathrm{val}}$ produced by Index Generator ({\color{blue}\S\ref{sec:indexing}}); and any subject-matter-expert metadata supplied with the benchmark. The output is the final focused schema $\mathit{Focused}_{\mathit{NLQ}}$.

%\algname{} 
The idea of the step is to generate preliminary SQL queries using an LLM based on three different prompts having:  $\mathit{Focused}^{\prime}_{\mathit{NLQ}}$, 
full schema of $\mathcal{D}_{\mathrm{target}}$, and $\mathit{Profile}_{T_i.A_j}$ with schema information.
Then, based on preliminary SQL queries, this step identifies any missing columns or tables and adds to the preliminary focused schema, resulting in the final $\mathit{Focused}_{\mathit{NLQ}}$.\footnote{\scriptsize 
~\cite{shkapenyuk2025automatic,cao2024rsl} inspired the process of schema linking.}

\subsection{\textsc{Phase~3}: SQL Generation}
\label{sec:sql-generation}
%As discussed in {\color{blue}\S\ref{sec:overview}}, 
\textsc{Phase~3} consists of two components: (\textit{i})~\emph{Note incorporator} ({\color{blue}\S\ref{sec:note-injection}}), which selects the question-relevant disambiguation notes $\mathcal{N}^{\star}_{\mathit{NLQ}}$; and 
(\textit{ii})~\emph{SQL generator}, which produces candidate SQL queries using the focused schema $\mathit{Focused}_{\mathit{NLQ}}$ produced by \textsc{Phase~2} and the selected notes. We describe each component below.

\subsubsection{\textbf{Note Incorporator}}
\label{sec:note-injection}
\emph{\textbf{Purpose.}} 
\algname{} creates a per-database note set $\mathcal{N}(\mathcal{D}_{\mathrm{target}})$ for ambiguous column pairs in \textsc{Step~4} of Deep Schema Explorator in \textsc{Phase~1} ({\color{blue}\S\ref{sec:deep-exploration}}).
However, during SQL generation, only a small subset of these notes is relevant to $\mathit{NLQ}$, since injecting all notes would add noise and may confuse the SQL-generating LLM. 
Note Incorporator, thus, selects only the notes needed for the current question.

\noindent
\emph{\textbf{Working.}} Given $\mathit{NLQ}$ over target database $\mathcal{D}_{\mathrm{target}}$, \algname{} takes the focused schema $\mathit{Focused}_{\mathit{NLQ}}$ produced by \textsc{Phase~2} and the database note set $\mathcal{N}(\mathcal{D}_{\mathrm{target}})$ produced by Deep Schema Explorator in \textsc{Phase~1} ({\color{blue}\S\ref{sec:deep-exploration}}).

It first filters $\mathcal{N}(\mathcal{D}_{\mathrm{target}})$ to candidate notes $\mathcal{N}_{\mathit{NLQ}}$. A note $\mathit{Note}_{\langle A_i,A_{\!j}\rangle}$ is retained as a candidate when at least one of its columns, $A_i$ or $A_j$, appears in $\mathit{Focused}_{\mathit{NLQ}}$. %This keeps notes that could matter for the columns currently under consideration. 
~
Then, this step asks an LLM to inspect $\mathit{NLQ}$, $\mathit{Focused}_{\mathit{NLQ}}$, and $\mathcal{N}_{\mathit{NLQ}}$ and decide which candidate notes are relevant to the question for preventing the SQL generator from choosing the wrong ambiguous column:
\[
  \mathcal{N}^{\star}_{\mathit{NLQ}}
  =
  \mathrm{LLM}\!\left(\mathit{NLQ},\mathit{Focused}_{\mathit{NLQ}},\mathcal{N}_{\mathit{NLQ}}\right)
  \subseteq
  \mathcal{N}_{\mathit{NLQ}}.
\]

The LLM labels each candidate note as relevant or irrelevant to $\mathit{NLQ}$. Notes labeled irrelevant are removed from the candidate-note set -- note that this operation does not remove any column from $\mathit{Focused}_{\mathit{NLQ}}$. The notes labeled `relevant' make up the selected note set $\mathcal{N}^{\star}_{\mathit{NLQ}}$ and are retained for prompt generation.

Furthermore, this step may expand $\mathit{Focused}_{\mathit{NLQ}}$ by adding new columns. Note that a note becomes a candidate whenever at least one of its two associated columns is already present in $\mathit{Focused}_{\mathit{NLQ}}$. 
However, a candidate note may contain another column that is not yet included in $\mathit{Focused}_{\mathit{NLQ}}$. In this case, the missing column is added to $\mathit{Focused}_{\mathit{NLQ}}$. %; that is, if $A_i$ (or $A_j$) is present but $A_j$ (or $A_i$) is absent, then $A_j$ (or $A_i$) is added to $\mathit{Focused}_{\mathit{NLQ}}$.
Importantly, Note Incorporator only expands $\mathit{Focused}_{\mathit{NLQ}}$ and never removes any existing column. 
If $\mathcal{N}^{\star}_{\mathit{NLQ}}=\varnothing$, no note is incorporated, and $\mathit{Focused}_{\mathit{NLQ}}$ remains unchanged.

% Furthermore, this step may add new columns to $\mathit{Focused}_{\mathit{NLQ}}$. Note that a note becomes a candidate when at least one of its two columns is present in $\mathit{Focused}_{\mathit{NLQ}}$; however, the selected note may contain one column that is absent in $\mathit{Focused}_{\mathit{NLQ}}$. 
% Such an absent column of the candidate note is added to $\mathit{Focused}_{\mathit{NLQ}}$, \textit{i}.\textit{e}., 
% if $A_i$ (or $A_{\!j}$) is present but $A_j$ (or $A_i$)  is absent,  $A_j$ (or $A_i$) is added to $\mathit{Focused}_{\mathit{NLQ}}$. Important to note that %Thus, 
% Note Incorporator 
% never removes an existing column from $\mathit{Focused}_{\mathit{NLQ}}$. 
% If $\mathcal{N}^{\star}_{\mathit{NLQ}}=\varnothing$, no note is incorporated and the focused schema remains unchanged.

%For each selected note $\mathit{Note}_{\langle A_i,A_j\rangle}\in\mathcal{N}^{\star}_{\mathit{NLQ}}$, whatever column is not this step updates $\mathit{Focused}_{\mathit{NLQ}}$, as follows: 

%may add a missing column identified by a selected note, but it 

\emph{\textbf{Example.}} For the running example in \textcolor{blue}{\S\ref{subsection:challenges_and_our_solutions}}, a question about a patient's final diagnosis may initially include \textsf{Examination.Diagnosis} in $\mathit{Focused}_{\mathit{NLQ}}$ while omitting \textsf{Patient.Diagnosis}. Because one column in the pair is present, the note comparing the two diagnosis columns becomes a candidate. The LLM selects this note as relevant because it explains that \textsf{Patient.Diagnosis} is patient-level while \textsf{Examination.Diagnosis} is exam-level. \algname{} then adds the missing \textsf{Patient.Diagnosis} column before SQL generation.

\subsubsection{\textbf{SQL Generator}}
\label{sec:sql-generator}
\emph{\textbf{Purpose.}} Different SQL questions may benefit from different reasoning processes, so relying on one generation path can cause all candidates to share the same failure. \algname{} uses complementary generation paths to produce diverse candidates and increase the likelihood that at least one captures the intended SQL. %\footnote{\scriptsize 
(The multi-path motivation is similar to~\cite{pourreza2024chase}.)

\noindent
\emph{\textbf{Working.}} The inputs are $\mathit{NLQ}$; the final focused schema $\mathit{Focused}_{\mathit{NLQ}}$ produced in \textsc{Step~2} and the matched database values retrieved in \textsc{Step~1} of \textsc{Phase~2} ({\color{blue}\S\ref{sec:schema-linking}}); the disambiguation notes $\mathcal{N}^{\star}_{\mathit{NLQ}}$ selected by Note Incorporator ({\color{blue}\S\ref{sec:note-injection}}); and any subject-matter-expert evidence supplied with the question in the benchmark. The output is a pool of candidate SQL queries, each denoted by $\mathit{candidate}_{\mathrm{SQL}}$, which is provided to SQL Correction and Selection in \textsc{Phase~4} ({\color{blue}\S\ref{sec:sql-correction-selection}}).

\algname{} provides these inputs to three generation paths. It combines a new dependency-tree-based generation strategy with two established strategies from prior work, as described below.

\begin{figure}[!t]%{r}{0.2\textwidth}
%\BBB\BB
    \centering
    \includegraphics[scale=0.3]{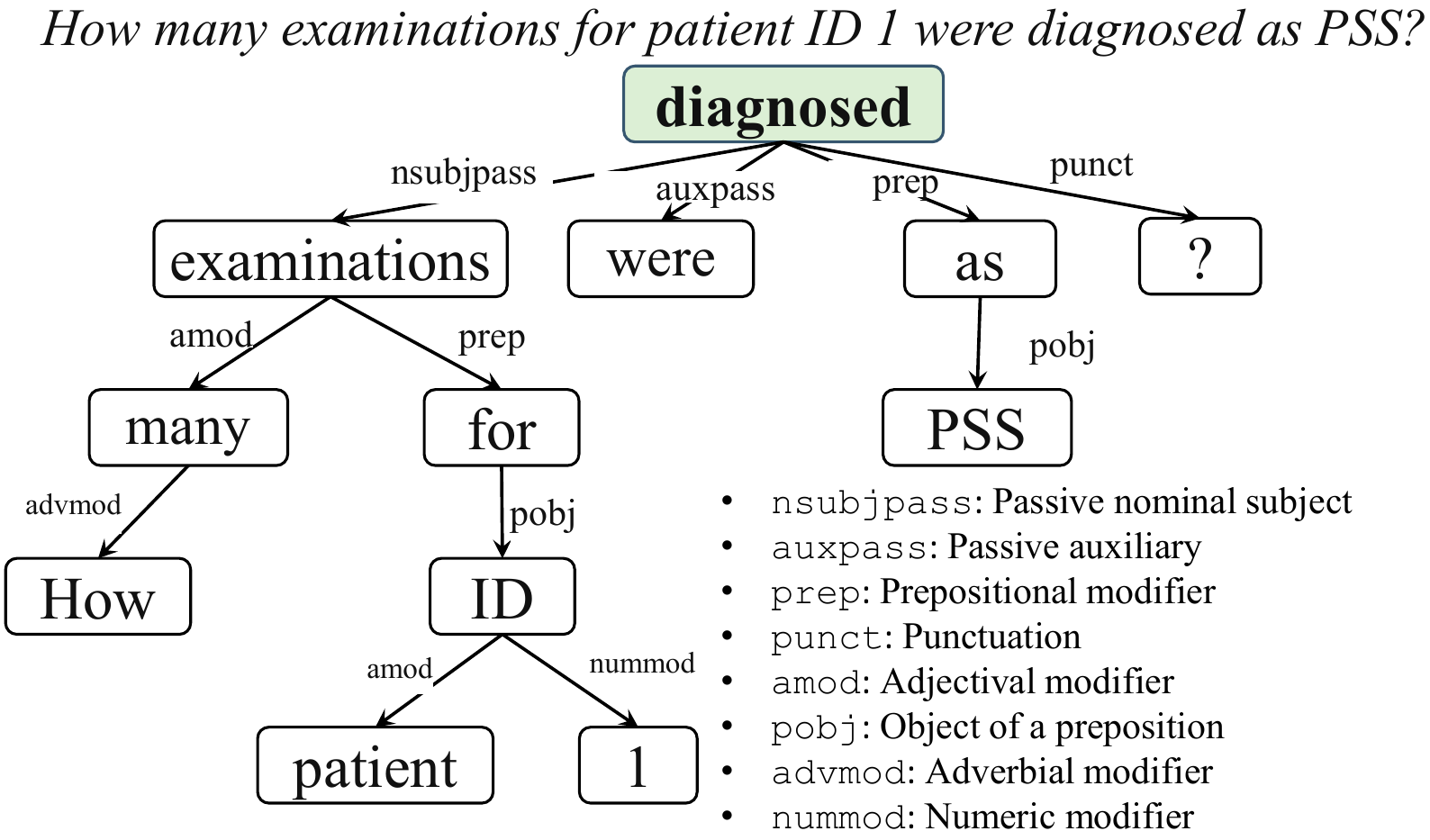}
    %\BBB%\BBB%\BBB
    \caption{Dependency-tree example.}% {\scriptsize \textnormal{Each node represents a question token, and each labeled edge represents the dependency between two tokens.}}}
     \label{fig:dependency-tree-generation}
     %\BBB\BB
\end{figure}

{
\textbf{\emph{Dependency-tree-based intermediate-representation generation}} captures the relationships among NLQ elements and uses them to improve SQL generation.  
To do so, a dependency tree is first computed from the $\mathit{NLQ}$ using a dependency parser~\cite{Honnibal2020}.

Each node of the dependency tree is a token of $\mathit{NLQ}$, and each directed, labeled edge states how one token modifies or depends on another; see {\color{blue}Figure~\ref{fig:dependency-tree-generation}}. 
Then, \algname{} traverses the dependency tree and deterministically extracts SQL-relevant features using eleven semantic units: \emph{entity}, \emph{attribute}, \emph{literal}, \emph{aggregate}, \emph{comparison}, \emph{negation}, \emph{conjunction}, \emph{ordering}, \emph{group}, \emph{limit}, and \emph{clause}. Each unit is derived deterministically using syntactic patterns over the tree and anchored to the specific span of $\mathit{NLQ}$ that triggers it. For example, in {\color{blue}Figure~\ref{fig:dependency-tree-generation}}, \textsf{`examinations'}, attached to \textsf{diagnosed} by \textsf{nsubjpass}, is extracted as an entity; a noun governed by a preposition is treated as a possible filter; 
\emph{`how many'} produces an aggregate unit with \textsc{count}; and \emph{at least} produces a comparison unit with $\geq$.
Then, \algname{} grounds each unit against $\mathit{Focused}_{\mathit{NLQ}}$ and the matched database values from \textsc{Phase~2} ({\color{blue}\S\ref{sec:schema-linking}}), resolving it to an actual table, column, or database literal, and assembles the grounded units into a structured intermediate representation. 
To generate $\mathit{candidate}_{\mathrm{SQL}}$, LLM receives $\mathit{NLQ}$, $\mathit{Focused}_{\mathit{NLQ}}$, the matched values, the selected disambiguation notes, SME metadata, and instructions to preserve the intermediate representation when constructing SQL. 
Then, LLM generates a candidate SQL that is checked against the intermediate representation to confirm that every component is present and no schema element outside $\mathit{Focused}_{\mathit{NLQ}}$ was introduced; a query that diverges is returned to LLM for revision.}

% \begin{figure}[t]
%     \centering
%     \includegraphics[
%         width=0.8\columnwidth
%     ]{dependency_tree_example.pdf}
%     \caption{Dependency-tree example. Each node represents a question token, and each labeled edge represents the dependency between two tokens.}
%     \label{fig:dependency-tree-generation}
% \end{figure}

\textbf{\emph{Few-shot in-context-learning generation}} uses similar training examples to demonstrate relevant question-to-SQL mappings. It retrieves structurally similar examples from the training set and provides them to the LLM as demonstrations.

\textbf{\emph{Divide-and-conquer generation}} makes complex questions more manageable by breaking them into simpler sub-questions. It generates a partial SQL solution for each sub-question and composes them into a complete query.\footnote{\scriptsize \cite{gao2023text,xie2025opensearch} used few-shot in-context learning. \cite{pourreza2024chase,li2026deepeye} used divide-and-conquer generation.}

\subsection{\textsc{Phase~4}: SQL Correction and Selection}
\label{sec:sql-correction-selection}
%As discussed in {\color{blue}\S\ref{sec:overview}}, 
\textsc{Phase~4} consists of two components: 
(\textit{i})~\emph{correction}, which repairs candidate SQLs using deterministic checks, execution feedback, and the correction rules created by Rule Creator ({\color{blue}\S\ref{sec:rule-creation}}); and 
(\textit{ii})~\emph{selection}, which chooses the most reliable candidate from the corrected pool. We describe each component below.

\subsubsection{\textbf{Correction}}
\label{sec:correction}
\emph{\textbf{Purpose.}} LLMs may hallucinate schema elements, produce malformed SQL, or generate queries that fail during execution. Correction detects and repairs these problems before the final SQL selection.

\noindent
\emph{\textbf{Working.}} \algname{} corrects each candidate in four steps. Step~1 creates a diagnostic report from deterministic syntax checking and execution feedback. Step~2 uses an LLM to revise the SQL using that report. Step~3 decides whether rule-based correction is needed and selects any relevant rules from Rule Creator in \textsc{Phase~1} ({\color{blue}\S\ref{sec:rule-creation}}). Step~4 applies the selected rules, when any are selected, to produce the corrected candidate. The steps are described below.

\smallskip
\noindent\textbf{\textsc{Step~1: Diagnostic report creation.}}
\emph{Purpose.} This step finds syntax and execution problems caused by malformed clauses, hallucinated identifiers, unsupported syntax, or incorrect query logic.

\emph{Method.} The inputs are a $\mathit{candidate}_{\mathrm{SQL}}$ produced by SQL Generator in \textsc{Phase~3} ({\color{blue}\S\ref{sec:sql-generator}}) and the target database $\mathcal{D}_{\mathrm{target}}$, and the output is a diagnostic report for that candidate. 
~
This step first checks the candidate for syntax and parse errors using SQLGlot~\cite{sqlglot}  which parses SQL into a dialect-aware abstract syntax tree and exposes malformed clauses or unsupported syntax. Then, this step executes the candidate SQL and records execution feedback, such as runtime errors, empty results, or suspicious outputs. 
These syntax and execution signals are combined into a diagnostic report.

\smallskip
\noindent\textbf{\textsc{Step~2: LLM revision.}}
\emph{Purpose.} This step repairs candidates using the diagnostic evidence from Step~1.

\emph{Method.} The inputs are $\mathit{candidate}_{\mathrm{SQL}}$ and its diagnostic report from \textsc{Step~1}, together with $\mathit{NLQ}$, $\mathit{Focused}_{\mathit{NLQ}}$, the selected disambiguation notes $\mathcal{N}^{\star}_{\mathit{NLQ}}$ from Note Incorporator ({\color{blue}\S\ref{sec:note-injection}}), and any subject-matter-expert evidence supplied with the question in the benchmark, and the output is the revised SQL $\mathit{revise}_{\mathrm{SQL}}$. This step invokes an LLM in revision mode with these inputs so that it can repair syntax errors, execution failures, or suspicious outputs while preserving the intent of $\mathit{NLQ}$.

\smallskip
\noindent\textbf{\textsc{Step~3: Rule relevance selection.}}
\emph{Purpose.} This step determines whether the revised SQL still needs rule-based correction.

\emph{Method.} The inputs are $\mathit{revise}_{\mathrm{SQL}}$, $\mathit{NLQ}$, and the rule set $\mathit{Rule}$ produced by Rule Creator in \textsc{Phase~1} ({\color{blue}\S\ref{sec:rule-creation}}), and the output is a (possibly empty) set of selected rules. 
~
This step parses and re-executes $\mathit{revise}_{\mathrm{SQL}}$ to generate an updated diagnostic report. It then uses an LLM to decide whether rule-based correction is still needed and, if so, which $\mathit{rule}_i\in\mathit{Rule}$ apply, using the question, revised SQL, updated diagnostic report, and rule descriptions as context. If rule-based correction is not needed, no rule is selected.

\smallskip
\noindent\textbf{\textsc{Step~4: Rule-guided correction.}}
\emph{Purpose.} This step applies only the selected rules from the previous step to fix recurring SQL-generation errors.

\emph{Method.} The inputs are $\mathit{revise}_{\mathrm{SQL}}$ and the selected rules from \textsc{Step~3}, and the output is a corrected SQL candidate. 

If one or more rules are selected, this step uses them to correct the SQL. Each $\mathit{rule}_i$ provides a \textit{gist}, \textit{bad-pattern}, \textit{correct-pattern}, and \textit{fix} instruction (see \textsc{Step}~4 of Rule Creator in {\color{blue}\S\ref{sec:rule-creation}}), which guide the rewrite of $\mathit{revise}_{\mathrm{SQL}}$. If no rule is selected, the revised SQL is kept unchanged. Applying this step to every revised candidate produces the corrected SQL candidate set $\mathit{correct}_{\mathrm{SQL}}$.

\subsubsection{\textbf{Selection}}
\label{sec:selection}
\emph{\textbf{Purpose.}} Since correction produces multiple SQL candidates that may return different results, Selection chooses the most reliable candidate as the final SQL.

\noindent
\emph{\textbf{Working.}} The input is the corrected SQL candidate set $\mathit{correct}_{\mathrm{SQL}}$ from Correction ({\color{blue}\S\ref{sec:correction}}), where each candidate has passed through revision and rule-guided correction, and  the output is $\mathit{final}_{\mathrm{SQL}}$. 

\algname{} first executes every corrected candidate and clusters candidates that produce identical result sets. The clusters are ranked by size, and each top cluster, denoted by $\mathit{Cluster}_i$, is represented by one SQL candidate $s_i$.
Then, \algname{} assigns $s_i$ an execution-confidence score equal to the fraction of corrected candidates that fall in its cluster, as:
$
  \mathit{Conf}(s_i)
  =
  {|\mathit{Cluster}_i|}/{|\mathit{correct}_{\mathrm{SQL}}|}
$.

If the execution confidence of the top-ranked representative $s_i$ exceeds a threshold, \algname{} treats the execution agreement as strong enough and returns that representative as $\mathit{final}_{\mathrm{SQL}}$. Otherwise, \algname{} selects representatives from the top-$K$ clusters and asks an LLM to adjudicate them pairwise using $\mathit{NLQ}$, $\mathit{Focused}_{\mathit{NLQ}}$, the selected disambiguation notes $\mathcal{N}^{\star}_{\mathit{NLQ}}$ from Note Incorporator ({\color{blue}\S\ref{sec:note-injection}}), any subject-matter-expert evidence supplied with the question in the benchmark, the candidate SQLs, and their execution-confidence scores. During comparison, candidates supported by more SQLs are favored unless their logic conflicts with the NL question. \algname{} then counts how often each candidate is preferred across the pairwise comparisons, combines this comparison support with its execution confidence, and returns the highest-scoring candidate as $\mathit{final}_{\mathrm{SQL}}$.

\section{Experimental Evaluation}
\label{sec:Experimental Evaluation}

This section experimentally compares \algname{} against existing non-fine-tuning-based Text-to-SQL methods. 
%\algname{} code is available at: \url{https://github.com/SecretDeB/DexterSQL}
%We first provide details of the entire experimental setup and then the results. 

% In this section, we evaluate the following: 

% 1.

% 2.

% 3.

\subsection{Dataset, Setup, and Baselines}
\label{subsec:Dataset, Setup, and Baselines}

% \subsubsection{
\smallskip
\noindent
\textbf{Datasets.}
We use two datasets:
(\textit{i})~\emph{\textbf{BIRD~\cite{li2023can}}} %is a large-scale Text-to-SQL benchmark designed to capture realistic database use cases. The complete benchmark 
contains 12,751 NLQ-SQL pairs over 95 databases from 37 domains. In our experiments, 1{,}534 NLQ-SQL pairs over 11 databases in the full \emph{BIRD-Dev} form $\mathcal{D}_{\mathrm{target}}$, and 9,428 NLQ-SQL pairs over 69 databases in BIRD-Train form $\mathcal{D}_{\mathrm{train}}$, of which randomly selected $\approx$~3,000 NLQ-SQL pairs form $\mathcal{D}_{\mathrm{train}}$ for % and are used only by Rule Creator .
Rule Creator ({\color{blue}\S\ref{sec:rule-creation}}). %, \algname{} samples 
~
(\textit{ii})~\emph{\textbf{Spider~\cite{yu2018spider}}} contains 10,181 NLQ-SQL pairs over 200 databases from 138 domains. In our experiments, 2,147 NLQ-SQL pairs  over 40 databases in \emph{Spider-Test} form $\mathcal{D}_{\mathrm{target}}$, and 7,000 NLQ-SQL pairs  over 140 databases in Spider-Train form $\mathcal{D}_{\mathrm{train}}$, of which $\approx$3,000 are used only by Rule Creator. % ({\color{blue}\S\ref{sec:rule-creation}}). 
% (\textit{ii})~\emph{\textbf{Spider~\cite{yu2018spider}}} contains 10,181 questions and 5,693 unique SQL queries over 200 databases from 138 domains across all splits. In our experiments, 2,147 questions over 40 databases in \emph{Spider-Test} form $\mathcal{D}_{\mathrm{target}}$, while the 7,000 questions over 140 databases in Spider-Train form $\mathcal{D}_{\mathrm{train}}$ and $\approx$3,000 are used only by Rule Creator ({\color{blue}\S\ref{sec:rule-creation}}). 

%\smallskip
\noindent
\textbf{Evaluation Metrics.}
\label{sec:evaluation_metrics}
We use the following four metrics: % in different experimental results:
~
\begin{enumerate}[nolistsep,noitemsep,leftmargin=0pt,]
\item
{\emph{Execution Accuracy (EX)}}: considers a generated SQL as correct when its execution result matches that of the gold SQL. 
EX is used to evaluate SQL-generation performance in  Exps.~1, 2, 4, 5 and~6. 
\item {\emph{Upper-bound Execution Accuracy (UB-EX)}}: finds the percentage of questions for which at least one generated candidate has the same execution result as the gold SQL (assuming an oracle always selects the correct candidate). 
UB-EX measures the maximum accuracy achievable from the candidate pool. Exp.~4 uses UB-EX. % also.
%
%
%\item
%
%
\item 
{\emph{Recall and Precision}}: Recall measures the fraction of columns referenced by the gold SQL that are retained in the focused schema. Precision measures the fraction of focused-schema columns referenced by the gold SQL. Exp.~3 uses recall and precision.

\item
\emph{Valid Efficiency Score (VES)}~\cite{li2023can}: measures the execution efficiency of generated SQL queries that return the same results as the gold SQL. For $N$ questions, it is computed as:
\[
\mathrm{VES}
=
\frac{1}{N}\sum_{i=1}^{N}
\mathrm{EX}\!\left(\mathit{gold}_{\mathrm{SQL}}^{(\textit{i})},\mathit{final}_{\mathrm{SQL}}^{(\textit{i})}\right)
\sqrt{
\frac{t_{\mathrm{gold}}^{(\textit{i})}}
     {t_{\mathrm{final}}^{(\textit{i})}}
}.
\]
Here, $\mathit{gold}_{\mathrm{SQL}}^{(\textit{i})}$ and $\mathit{final}_{\mathrm{SQL}}^{(\textit{i})}$ denote the gold SQL and final SQL for the $i$th NLQ query, respectively. Here, $\mathrm{EX}(\cdot,\cdot)$ denotes whether the execution results of the two SQL queries match, as defined in EX Accuracy above. And, $t_{\mathrm{gold}}^{(\textit{i})}$ and $t_{\mathrm{final}}^{(\textit{i})}$ denote the execution times of the $i$th SQL query. Exp.~7 uses VES.

\end{enumerate}
%, 2, 3, 4, 5, and~6.
%
%
%
%
%\item 
%We also use 
%additionally reports 
~
\noindent
\textbf{Setup.} Experiments were conducted on an HPC cluster running Linux with NVIDIA A100 80GB GPUs. To validate the effectiveness of \algname{}, we use three distinct models: the open-weight GPT-OSS-120B model and the closed-weight GPT-4o and GPT-5.2 models. For indexing  in \textsc{Phase~1} ({\color{blue}\S\ref{sec:indexing}}), we use Qwen3-Embedding-0.6B. We generate four candidates per SQL generation method (12 total) ({\color{blue}\S\ref{sec:sql-generator}}) and set the Selection threshold to $\mathit{Conf}(s_i)>0.6$ ({\color{blue}\S\ref{sec:selection}}).

%We evaluate \algname{} with GPT-4o and GPT-5.2 as frontier-model backbones. In addition, we evaluate GPT-OSS-120B as an open-weight backbone to test whether \algname{} also improves accuracy in a lower-cost local setting. Although GPT-OSS-120B has roughly 117B total parameters, its mixture-of-experts architecture activates only about 5.1B parameters per token, making it a relatively small active-parameter open-weight setting compared with frontier models. For semantic value retrieval, \algname{} uses Qwen3-Embedding-0.6B and retrieves the top-5 similar values for each \textsf{TEXT} column. 

%\subsubsection{
%\smallskip
\noindent
\textbf{Baselines.} 
%\label{sec:experiment-baselines}
\algname{} is a non-fine-tuning framework, thus, we compare it against {\bf 10 prompt-based}, training-free Text-to-SQL methods, as: %. These baselines are: 
DAIL-SQL~\cite{gao2023text}, 
DIN-SQL~\cite{pourreza2023din}, C3~\cite{dong2023c3}, 
RSL-SQL~\cite{cao2024rsl}, 
OpenSearch-SQL~\cite{xie2025opensearch}, Rethinking Schema Linking~\cite{nahid2026rethinking}, 
Alpha-SQL~\cite{li2025alpha}, AutoLink~\cite{wang2026autolink}, 
DeepEye-SQL~\cite{li2026deepeye}, and ApexSQL~\cite{cao2026apex}.

\begin{table}[t]
%\BBB%\BBB
\centering
\small
%\scriptsize
% \setlength{\tabcolsep}{4pt}
% \resizebox{\columnwidth}{!}{%
\caption{Exp.~1: Execution accuracy of non-fine-tuning methods using % on Spider-Test and BIRD-Dev. 
%All methods are run locally with 
GPT-OSS-120B.
}
\begin{tabular}{|l|c|c|}\hline
\textbf{Method} & \textbf{Spider-Test} & \textbf{BIRD-Dev} \\ \hline
DAIL-SQL~\cite{gao2023text} & 74.0 & 51.8 \\ \hline
C3~\cite{dong2023c3} & 69.3 & 53.8 \\ \hline

Rethinking Schema Linking~\cite{nahid2026rethinking} & 72.2 & 54.6 \\ \hline

DIN-SQL~\cite{pourreza2023din} & 70.0 & 56.2 \\ \hline
AutoLink~\cite{wang2026autolink} & 77.5 & 57.4 \\ \hline
OpenSearch-SQL~\cite{xie2025opensearch} & 67.8 & 58.2 \\ \hline
RSL-SQL~\cite{cao2024rsl} & 74.6 & 59.3 \\ \hline
Alpha-SQL~\cite{li2025alpha} & 80.3 & 62.8 \\ \hline
APEX-SQL~\cite{cao2026apex} & 79.1 & 64.2 \\ \hline
DeepEye-SQL~\cite{li2026deepeye} & 81.9 & 64.9 \\ \hline
\rowcolor[HTML]{C1FD8E} \algname{} & 85.6 & 70.4 \\ \hline
\end{tabular}%
% }

\label{tab:gpt-oss-results}
%\BBB%\BBB\BB
\end{table}

\subsection{Experimental Results}
\label{sec:experimental-evaluation}

In this section, we evaluate the following:
\begin{enumerate}[leftmargin=0.01in, label=(\arabic*), noitemsep, nolistsep]
    \item SQL-generation accuracy with a open-weight model, Exp.~1 ({\color{blue}\S\ref{sec:experiment-sql-generation-performance-I}}).

    \item SQL-generation accuracy on closed-weight models, Exp.~2 ({\color{blue}\S\ref{sec:experiment-sql-generation-performance-II}}).

    \item Recall and precision of the final focused schema, Exp.~3 ({\color{blue}\S\ref{sec:experiment-schema-linking}}).
    
    \item Individual and combined contributions of SQL generation method/paths, Exp.~4 ({\color{blue}\S\ref{sec:experiment-generator-analysis}}).
    
    \item Effects of the confidence threshold and SQL selection strategies, Exp.~5 ({\color{blue}\S\ref{sec:experiment-confidence-selection}}).
    
    \item Contributions of Deep Schema Explorator, dependency-tree-based generation, and Rule Creator-based correction, Exp.~6 ({\color{blue}\S\ref{sec:ablation-study}}).
\item Execution efficiency of correctly generated SQL queries, Exp.~7 ({\color{blue}\S\ref{sec:experiment-ves}}). 
%    \item SQL-generation accuracy across different SQL characteristics, Exp.~8 ({\color{blue}\S\ref{sec:experiment-sql-characteristics}}).}
\end{enumerate}

\subsubsection{\textbf{Exp.~1: SQL-Generation Accuracy with Open-Weight model.}}
\label{sec:experiment-sql-generation-performance-I}

%The goal of t
This experiment finds  \algname{} performance using open-weight LLM model; see {\color{blue}Table~\ref{tab:gpt-oss-results}}. We executed all baseline methods, with available implementations, ourselves  locally using GPT-OSS-120B, which activates about 5.1B parameters per token. %, and compare their accuracies with \algname{}. 
% shows the results.

\smallskip
\begin{mdframed}[
  linewidth=0.6pt,
  backgroundcolor=yellow!8,
  innerleftmargin=2pt,
  innerrightmargin=2pt,
  innertopmargin=2pt,
  innerbottommargin=2pt
]
\noindent\textbf{Findings.} 
\algname{} achieves the highest accuracy on both benchmarks, reaching 85.6\% on Spider-Test and 70.4\% on BIRD-Dev. On BIRD-Dev, the closest baseline is DeepEye-SQL at 64.9\%, compared with 70.4\% for \algname{}, an improvement of 5.5\% by our system, and the same holds on Spider-Test --- DeepEye-SQL is again the closest baseline at 81.9\%, while \algname{} reaches 85.6\%, showing an improvement of 3.7\%.
\end{mdframed}

\begin{table}[t]
\centering
\small %criptsize
% \setlength{\tabcolsep}{3pt}
% \resizebox{\columnwidth}{!}{%
\caption{Exp.~2: Execution accuracy (EX) on BIRD-Dev using GPT-4 or GPT-4o. S: Simple. M: Moderate. C: Challenging. }%$^{*}$ denotes local reproduction.}
\begin{tabular}{|l|l|c|c|c|c|}\hline
\textbf{Method} & \textbf{LLM} & \textbf{S} & \textbf{M} & \textbf{C} & \textbf{Total} \\
       % &     & (925) & (464) & (145) & (1534) \\ 
       \hline
C3~\cite{dong2023c3} & GPT-4 & 58.9 & 38.5 & 31.9 & 50.2 \\ \hline
DAIL-SQL~\cite{gao2023text} & GPT-4 & 62.5 & 43.2 & 37.5 & 54.3 \\ \hline
TA-SQL~\cite{qu2024before} & GPT-4 & 63.1 & 48.6 & 36.1 & 56.2 \\ \hline
MAG-SQL~\cite{xie2024mag} & GPT-4 & 65.9 & 46.2 & 41.0 & 57.6 \\ \hline
SuperSQL~\cite{li2024dawn} & GPT-4 & 66.9 & 46.5 & 43.8 & 58.5 \\ \hline
MAC-SQL~\cite{wang2025mac} & GPT-4 & 65.7 & 52.7 & 40.3 & 59.4 \\ \hline
MCS-SQL~\cite{lee-etal-2025-mcs} & GPT-4 & 70.4 & 53.1 & 51.4 & 63.4 \\ \hline
RSL-SQL~\cite{cao2024rsl} & GPT-4o & 74.4 & 57.1 & 53.8 & 67.2 \\ \hline
OpenSearch-SQL~\cite{xie2025opensearch} & GPT-4o & -- & -- & -- & 69.3 \\ \hline
DeepEye-SQL~\cite{li2026deepeye}%$^{*}$
& GPT-4o & 75.7 & 61.6 & 59.3 & 69.9 \\ \hline
APEX-SQL~\cite{cao2026apex} & GPT-4o & 75.9 & 64.4 & 57.2 & 70.7 \\ \hline
\rowcolor[HTML]{C1FD8E} \algname{} & GPT-4o & 77.2 & 64.8 & 63.2 & 72.1 \\ \hline
\end{tabular}%
% }

\label{tab:bird-sql-generation}
%\BBB%\BBB\BB
\end{table}

\begin{table}[t]
\centering

\small %criptsize

% \setlength{\tabcolsep}{3pt}
% \resizebox{\columnwidth}{!}{%
\caption{Exp.~2: EX on BIRD-Dev using GPT-5.2.}
\begin{tabular}{|l|l|c|c|c|c|}\hline
\textbf{Method} & \textbf{LLM} & \textbf{S} & \textbf{M} & \textbf{C} & \textbf{Total} \\
       % &     & (925) & (464) & (145) & (1534) \\ 
       \hline
DeepEye-SQL~\cite{li2026deepeye} & GPT-5.2 & 75.1 & 61.0 & 58.7 & 69.3 \\ \hline
APEX-SQL~\cite{cao2026apex} & GPT-5.2 & 74.9 & 63.4 & 56.2 & 69.7 \\ \hline
\rowcolor[HTML]{C1FD8E} \algname{} & GPT-5.2 & 77.0 & 66.7 & 66.7 & 72.9\\ \hline
\end{tabular}%
% }
% All methods are run locally.}% {\color{red} comment for me --- Controlled GPT-5.2 comparison across BIRD-Dev difficulty levels. All methods are run locally}???}
%Controlled GPT-5.2 comparison across BIRD-Dev difficulty levels. All methods are run locally.}
\label{tab:gpt52-sql-generation}
%\BBB%\BBB\BB
\end{table}
%anik:add heading , on VES say that we are good

\begin{table}[!t]
%%\BBB%\BBB
\centering
\small
%%\scriptsize
% \setlength{\tabcolsep}{5pt}
\caption{Exp.~2: Comparison with non-fine-tuning methods using different frontier models on BIRD-Dev.}
\begin{tabular}
{|l|l|c|}\hline
\textbf{Method} & \textbf{LLM} & \textbf{BIRD-Dev EX (\%)} \\\hline
CHESS~\cite{talaei2024chess} & Gemini-1.5-Pro & 68.3 \\ \hline
DSR-SQL~\cite{hao2025text} & DeepSeek-V3.1 & 68.3 \\ \hline
AutoLink~\cite{wang2026autolink} & Gemini-1.5-Pro & 68.7 \\ \hline
\rowcolor[HTML]{C1FD8E} \algname{} & GPT-4o & 72.1 \\ \hline
\end{tabular}

\label{tab:diverse-backbone-sql-generation}
%\BBB%\BBB\BB
\end{table}

\subsubsection{\textbf{Exp.~2:  SQL-Generation Accuracy across Closed-Weight Models.}}
\label{sec:experiment-sql-generation-performance-II}

This experiment  compares SQL-generation accuracy of \algname{} against the baseline approaches using closed-weight models: 
% :  
% %We use 
 GPT-4, GPT-4o, 
% %in 
  and GPT-5.2. % in . database

\textbf{\emph{EX on GPT-4 and GPT-4o ({\color{blue}Table~\ref{tab:bird-sql-generation}})}: }
% {\color{red}\bf run locally on table names may be confusing. I guess run locally means. We ran the code ourselves? please clarify.}
{%\color{cyan} -- MURAT
%For this experiment, 
We included only the published works\footnote{\scriptsize Published in a conference or a journal only, or provided the code.} that reported %use the 
accuracy using GPT-4 and 4o.
Also, we selected \emph{only DeepEye-SQL here} and ran it ourselves locally on GPT-4o, since results using GPT-OSS-120B  in {\color{blue}Table~\ref{tab:gpt-oss-results}} show that DeepEye-SQL is a close competitor to \algname{}. 
% So, we executed DeepEye-SQL locally on GPT-4o ourselves.
% %in the respective papers for all methods except DeepEye-SQL. Since DeepEye-SQL does not report results with GPT-4o, we reproduce its results locally using GPT-4o.} 
On such closed-weight models, 
\algname{} also achieves the highest EX of 72.1\%, exceeding the strongest baseline (i.e., APEX-SQL) by 1.4\%.}

\textbf{\emph{EX on GPT-5.2 ({\color{blue}Table~\ref{tab:gpt52-sql-generation}})}:} 
For GPT-5.2, we select only APEX-SQL and DeepEye-SQL because the GPT-4/GPT-4o results in {\color{blue}Table~\ref{tab:bird-sql-generation}} show that they are the two closest competitors to \algname{}. {\color{blue}Table~\ref{tab:gpt52-sql-generation}} shows that \algname{} achieves 72.9\% EX, exceeding APEX-SQL at 69.7\% and DeepEye-SQL at 69.3\%. Thus, \algname{} outperforms the strongest baseline by 3.2 percentage points.\footnote{\scriptsize
Although the APEX-SQL paper reports 70.7\% with GPT-4o, running the same released APEX-SQL code with GPT-5.2 in our setting produces a lower accuracy of 69.7\%.}

\textbf{\emph{EX on other models.}}
Some systems report results with other LLMs. {\color{blue}Table~\ref{tab:diverse-backbone-sql-generation}} compares \algname{} with CHESS and AutoLink using Gemini-1.5-Pro and DSR-SQL using DeepSeek-V3.1, based on the results reported in the respective papers. The strongest of these systems is AutoLink at 68.7\%, which \algname{} exceeds by 3.4\%. 

\smallskip
\begin{mdframed}[
  linewidth=.6pt,
  backgroundcolor=yellow!8,
  innerleftmargin=2pt,
  innerrightmargin=2pt,
  innertopmargin=2pt,
  innerbottommargin=2pt
]
\noindent\textbf{Findings.} {\algname{} consistently outperforms existing non-fine-tuning Text-to-SQL systems across different  models.}
\end{mdframed}

\begin{table}[!t]
%\BBB%\BBB
\centering
\small
%\scriptsize
% \setlength{\tabcolsep}{8pt}
\caption{Exp.~3: Schema-linking recall \& precision. }% on BIRD-Dev.}
\begin{tabular}{|l|c|c|}\hline
\textbf{Method on BIRD-Dev} & \textbf{Recall (\%)} & \textbf{Precision (\%)} \\ \hline
ReFoRCE~\cite{deng2025reforce} & 51.29 & 68.05 \\ \hline
AutoLink~\cite{wang2026autolink} & 74.39 & 16.27 \\ \hline
DeepEye-SQL~\cite{li2026deepeye} & 95.40 & 49.5 \\ \hline
RSL-SQL~\cite{cao2024rsl} & 95.75 & 47.90 \\ \hline
APEX-SQL~\cite{cao2026apex} & 96.15 & 51.95 \\ \hline
\rowcolor[HTML]{C1FD8E} \algname{} & 97.09 & 72.26 \\ \hline
\end{tabular}

\label{tab:bird-schema-linking}
%\BBB%\BBB\B
\end{table}

\subsubsection{\textbf{Exp.~3: Evaluating Schema-Linking Accuracy.}}
\label{sec:experiment-schema-linking}

%The goal of 
This experiment %is to 
finds whether \algname{} retains the schema elements required for SQL generation while excluding irrelevant schema context more effectively than existing training-free schema-linking methods. 
 %on BIRD-Dev set. 
We evaluate schema linking on BIRD-Dev using recall and precision, as defined in %Evaluation Metrics (
{\color{blue}\S\ref{sec:evaluation_metrics}}, as in prior schema-linking evaluations~\cite{cao2026apex}.
~
For \algname{}, we measure recall and precision over the final focused schema used for SQL generation with GPT-OSS-120B. 
This schema is produced by the schema-linking phase 
({\color{blue}\S\ref{sec:schema-linking}}) and then enriched by  Note incorporator ({\color{blue}\S\ref{sec:note-injection}}), which uses the outputs of Deep Schema Explorator ({\color{blue}\S\ref{sec:deep-exploration}}). 
Recall that when a relevant ambiguity note identifies an ambiguous column missing from the initial focused schema, \algname{} adds that column before SQL generation --- recovering required ambiguous columns that schema linking alone may omit, while the \emph{note-relevance filtering limits unnecessary additions to preserve precision}.
~
{\color{blue}Table~\ref{tab:bird-schema-linking}} reports  recall and precision of \algname{} and five \emph{training-free schema-linking methods} on BIRD-Dev set.

\smallskip
\begin{mdframed}[
  linewidth=0.6pt,
  backgroundcolor=yellow!8,
  innerleftmargin=2pt,
  innerrightmargin=2pt,
  innertopmargin=2pt,
  innerbottommargin=2pt
]
\noindent\textbf{Findings.}
Among the baseline approaches with schema linking, \algname{} achieves the highest recall and precision, reaching 97.09\% and 72.26\%, respectively. It exceeds the strongest baseline, APEX-SQL, by 0.94\% points in recall, and the strongest precision baseline, ReFoRCE, by 4.21\% points, showing that \algname{} retains more columns required by the gold SQL while introducing less irrelevant schema context than the other approaches.
\end{mdframed}

\subsubsection{\textbf{Exp.~4: Evaluating Contributions of the SQL Generation Methods.}}
\label{sec:experiment-generator-analysis}
\begin{figure}[t]
%\BBB%\BBB
    \centering
    \includegraphics[scale=0.44]{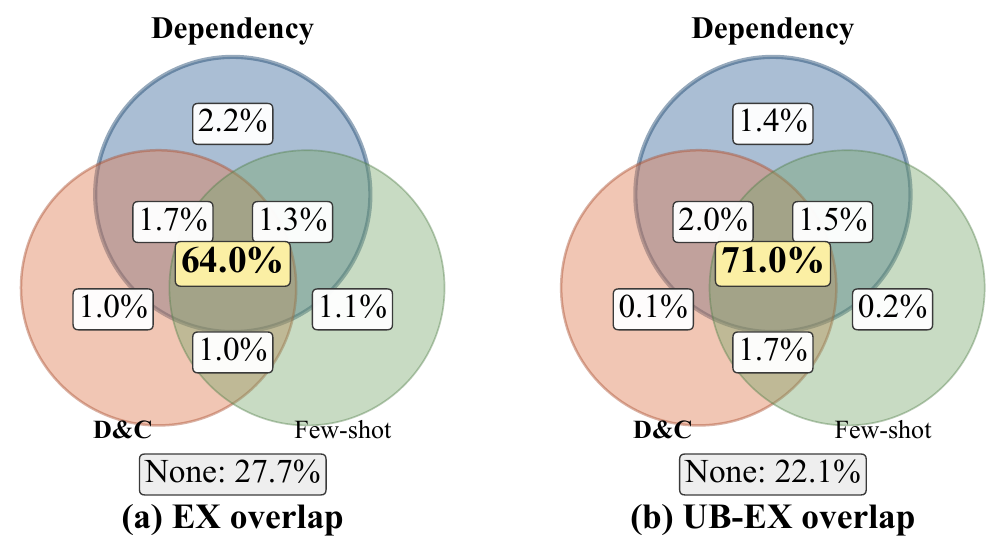}
      %\BBB%\BBB %\BBB
    \caption{Exp. 4: Correctness overlap among dependency-tree-based, few-shot in-context-learning, and divide-and-conquer SQL generation on BIRD-Dev using GPT-OSS-120B: (a) EX overlap and (b) UB-EX overlap.}
    \label{fig:generator-correctness-overlap}
    %\BBB\BB
\end{figure}

The goal of this experiment is to measure the impact of our dependency-tree-based SQL generation method ({\color{blue}\S\ref{sec:sql-generator}}) and to find the benefits of combining it with the established few-shot and divide-and-conquer methods. 
We evaluate each method individually, then all three together on BIRD-Dev with GPT-OSS-120B. All other pipeline components remain unchanged. %, so that the observed differences can be attributed to the active SQL generators.
For each configuration, we report EX and UB-EX (as defined in {\color{blue}\S\ref{subsec:Dataset, Setup, and Baselines}}).
{\color{blue}Table~\ref{tab:generator-analysis}} reports the individual and combined performance of the three SQL generation methods.
~ 
{\color{blue}Figure~\ref{fig:generator-correctness-overlap}} shows whether the methods succeed on the same or different questions. Each circle represents the questions answered correctly by one method, while overlapping regions represent questions answered correctly by multiple methods. In the EX analysis, all three methods answer 64.0\% of questions correctly, with another 8.3\% answered by only one or two methods. In the UB-EX analysis, all three methods cover 71.0\% of questions, with the remaining 6.9\% covered by only one or two methods.

\smallskip
\begin{mdframed}[
  linewidth=0.6pt,
  backgroundcolor=yellow!8,
  innerleftmargin=2pt,
  innerrightmargin=2pt,
  innertopmargin=2pt,
  innerbottommargin=2pt
]
\noindent\textbf{Findings.}
The dependency-tree-based method is the strongest individual SQL generator, achieving 69.2\% EX and 75.9\% UB-EX. Combining all three methods increases EX to 70.4\% and UB-EX to 77.9\%, showing that the few-shot and divide-and-conquer methods provide complementary candidates.
\end{mdframed}

%EX measures the accuracy of the SQL selected by the complete pipeline, while UB-EX measures the percentage of questions for which at least one generated candidate is execution-equivalent to the gold SQL under oracle selection.

\begin{table}[!t]
\centering
\small
%\scriptsize
% \setlength{\tabcolsep}{5pt}
\caption{Exp.~4: Individual and combined performance of the SQL generation paths  on BIRD-Dev using GPT-OSS-120B.}
\begin{tabular}
{|l|c|c|}\hline
\textbf{Active SQL generation paths} & \textbf{EX (\%)} & \textbf{UB-EX (\%)}

\\ \hline
% TODO: Replace these illustrative values with measured experimental results.
Dependency-tree-based & 69.2 & 75.9 \\ \hline
Divide-and-conquer & 67.7 & 74.8 \\ \hline
Few-shot in-context learning & 67.4 & 74.4 \\ \hline
\rowcolor[HTML]{C1FD8E} All three paths & 70.4 & 77.9 \\ \hline

\end{tabular}

\label{tab:generator-analysis}
%\BBB%\BBB\BB
\end{table}

% {\color{blue}Table~\ref{tab:generator-analysis}} shows that the dependency-tree-based method is the strongest individual generator, achieving 69.2\% EX and 75.9\% UB-EX. The other two methods achieve  lower EX (67.4\% for few-shot and 67.7\% for divide-and-conquer). Combining them with the dependency-tree-based method raises EX to 70.4\% and UB-EX to 77.9\%, gains of 1.2\% and 2.0\% over the strongest individual method. 

%Anik: mark green in tables
%Table should be same as 1st table.

\begin{figure}[!t]
    \centering
    \includegraphics[scale=0.7]{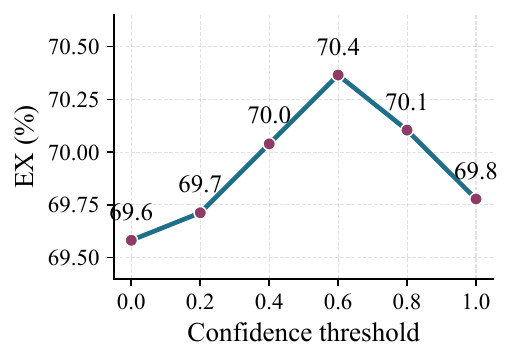}
    %\BBB%\BBB
    \caption{Execution accuracy across confidence shortcut thresholds on BIRD-Dev using GPT-OSS-120B.}
    \label{fig:confidence-threshold}
    %\BBB\B
\end{figure}

%Anik: mark green in tables
%Table should be same as 1st table.

\subsubsection{\textbf{Exp.~5: Evaluating Confidence-Aware SQL Selection.}}
\label{sec:experiment-confidence-selection}
The goal of this experiment is to find whether agreement among generated SQL candidates can identify when LLM review is needed during selection. In \textsc{Phase~4} ({\color{blue}\S\ref{sec:sql-correction-selection}}), \algname{} executes the corrected candidates and groups those producing the same result. 
If the fraction of candidates in the largest group exceeds a threshold, their agreement is considered sufficient, and the group's representative SQL is selected directly.
Otherwise, \algname{} uses an LLM to compare representatives from the top groups pairwise in a tournament-based selection. We evaluate this process on BIRD-Dev with GPT-OSS-120B by varying the threshold to find how effectively it separates cases that can rely on candidate agreement from those requiring LLM review.

We vary the threshold from 0 to 1 to control when LLM review is used. {\color{blue}Figure~\ref{fig:confidence-threshold}} shows EX rises from 69.6\% at threshold 0 to 70.4\% at 0.6, then falls to 69.8\% at threshold 1. These results show that selectively invoking LLM review when candidate agreement is insufficient improves SQL selection accuracy, with the highest EX achieved at a threshold of 0.6.
\smallskip
\begin{mdframed}[
  linewidth=0.6pt,
  backgroundcolor=yellow!8,
  innerleftmargin=2pt,
  innerrightmargin=2pt,
  innertopmargin=2pt,
  innerbottommargin=2pt
]
\noindent\textbf{Findings.}
At a threshold of 0.6, confidence-aware selection achieves the highest EX of 70.4\%. This confirms that using execution confidence to determine when LLM review is required improves SQL selection accuracy.
\end{mdframed}

\subsubsection{\textbf{Exp.~6: Evaluating Contributions of the Novel Components.}}
\label{sec:ablation-study}
This experiment measures the individual and combined contributions of \algname{}'s three novel components: Deep Schema Explorator ({\color{blue}\S\ref{sec:deep-exploration}}), Rule Creator-based correction ({\color{blue}\S\ref{sec:rule-creation}}), and dependency-tree-based intermediate-representation generation ({\color{blue}\S\ref{sec:sql-generator}}). We conduct the ablation on BIRD-Dev with GPT-OSS-120B by disabling each component separately and all three together. Disabling Deep Schema Explorator also disables Note Incorporator, while disabling Rule Creator disables rule-based correction; the other two generation methods remain active when dependency-tree-based generation is disabled.  {\color{blue}Table~\ref{tab:ablation-results}} presents the configurations in decreasing order of EX, from the full pipeline to the individual and combined ablations.
\smallskip
\begin{mdframed}[
  linewidth=0.6pt,
  backgroundcolor=yellow!8,
  innerleftmargin=2pt,
  innerrightmargin=2pt,
  innertopmargin=2pt,
  innerbottommargin=2pt
]
\noindent\textbf{Findings.}
The full \algname{} pipeline achieves 70.4\% EX. Removing Deep Schema Explorator, dependency-tree-based generation, or Rule Creator-based correction reduces EX to 68.4\%, 68.2\%, and 67.5\%, respectively, showing that each novel component contributes to accuracy. %Disabling all three components produces the largest decrease to 63.3\%, confirming that they provide complementary benefits when used together.
\end{mdframed}

\begin{table}[!t]
%%\BBB%%\BBB
\centering
\small
%\scriptsize
% \setlength{\tabcolsep}{3pt}
% \resizebox{\columnwidth}{!}{%
\caption{%Exp.~5: 
Ablation results on BIRD-Dev using OSS-120B.\protect\footnotemark}
\begin{tabular}{|l|c|}\hline
\textbf{Configuration} & \textbf{EX (\%)} \\ \hline
\rowcolor[HTML]{C1FD8E}
Full \algname{} pipeline & 70.4 \\ \hline
\rowcolor[HTML]{CCFCA6}
w/o Deep Schema Explorator~{\color{blue}(\S\ref{sec:deep-exploration})} & 68.4 \\ \hline
\rowcolor[HTML]{DAFBC0}
w/o Dependency-tree-based generation~{\color{blue}(\S\ref{sec:sql-generator})} & 68.2 \\ \hline
\rowcolor[HTML]{E2FBCD}
w/o Rule Creator-based correction~{\color{blue}(\S\ref{sec:rule-creation})} & 67.5 \\ \hline
\rowcolor[HTML]{F0FCE8}
w/o all three novel components & 63.8 \\ \hline
\end{tabular}
% }
\label{tab:ablation-results}
\end{table}
\footnotetext{\textnormal{``W/o all three novel components'' disables Deep Schema Explorator, dependency-tree-based intermediate-representation generation, and Rule Creator-based correction simultaneously. Darker green indicates higher EX.}}

 \subsubsection{\textbf{Exp.~7: Evaluating SQL Execution Efficiency.}}
\label{sec:experiment-ves}
The goal of this experiment is to determine the execution efficiency of the final SQL queries generated by \algname{}, measured using VES, as discussed in {\color{blue}\S\ref{sec:evaluation_metrics}}.

We evaluate \algname{} on BIRD-Dev with GPT-OSS-120B using VES and report results for the simple, moderate, and challenging subsets and the complete BIRD-Dev set. {\color{blue}Table~\ref{tab:bird-ves}} compares \algname{} with methods already included in our evaluation for which directly comparable BIRD-Dev VES results are available in prior work~\cite{li2024dawn}. \algname{} achieves the highest VES overall and across all difficulty levels, reaching 74.47, 54.17, and 52.04 on the simple, moderate, and challenging subsets, respectively, and 66.21 overall. It exceeds the next-highest overall result by 3.36 points, demonstrating stronger SQL execution efficiency.

\begin{table}[t]
\centering
%\scriptsize
\setlength{\tabcolsep}{3pt}
\begin{tabular}{|l|c|c|c|c|}\hline
\textbf{Method} & \textbf{Simple} & \textbf{Moderate} & \textbf{Challenging} & \textbf{All} \\ \hline
C3-SQL~\cite{dong2023c3} & 59.82 & 41.68 & 31.93 & 51.70 \\ \hline
DAIL-SQL~\cite{gao2023text} & 65.04 & 43.35 & 39.33 & 56.05 \\ \hline
SuperSQL~\cite{li2024dawn} & 69.75 & 50.55 & 49.08 & 61.99 \\ \hline
APEX-SQL~\cite{cao2026apex} & 69.06 & 53.47 & 50.18 & 62.56 \\ \hline
DeepEye-SQL~\cite{li2026deepeye} & 69.04 & 54.13 & 51.27 & 62.85 \\ \hline
\rowcolor[HTML]{C1FD8E} \algname{} & 74.47 & 54.17 & 52.04 & 66.21 \\ \hline
\end{tabular}
\caption{Exp.~7: Reward Valid Efficiency Score on BIRD-Dev across difficulty levels}
\label{tab:bird-ves}
%%\BBB%%\BBB
\end{table}

\section{Conclusion}

We developed  \algname{}, a non-fine-tuning Text-to-SQL system. % designed to improve database understanding, SQL generation, and correction without updating the underlying LLM parameters. 
\algname{} introduces three novel components:
\emph{Deep Schema Explorator} to uncover the different roles of ambiguous columns and create reusable disambiguation notes, 
\emph{Rule Creator} to convert recurring database-agnostic generation failures from training data into correction rules, and a 
\emph{dependency-tree-based SQL generator} that uses the question structure to construct an intermediate SQL representation.  Experimental results validate \algname{} effectiveness in generating SQL queries, showing higher accuracy on both 
 open-weight and closed-weight models.
 
\section*{Acknowledgments}
We are thankful to New Jersey Institute of Technology High Performance Computing (HPC) facility for providing the computational resources used in this work. We also thank Arda Ayna, Aruntej Thummepally, and Vineet Vora for their valuable assistance.

%\bibliographystyle{ACM-Reference-Format}
%\bibliography{references}

\end{document}